\documentclass[12pt,preprint]{emulateapj}
  
\usepackage{apjfonts}

\shorttitle{Properties of LAEs at z $\sim$ 0.3}
\shortauthors{Finkelstein et al.}

\newcommand{\sol}{$_{\odot}$}
\newcommand{\lya}{Ly$\alpha$}

\newcommand{\fig}[1]{Figure~\ref{#1}}
\def\arcs{\hbox{$^{\prime\prime}$}}

\begin{document}
\title{Dust Extinction and Metallicities of Star--Forming Lyman Alpha Emitting Galaxies at Low Redshift\altaffilmark{1}}

\author{Steven   L.    Finkelstein\altaffilmark{2,*}, Seth H. Cohen\altaffilmark{3}, John Moustakas\altaffilmark{4}, Sangeeta Malhotra\altaffilmark{3}, James E. Rhoads\altaffilmark{3}\\ and Casey Papovich\altaffilmark{2}}   
\altaffiltext{1}{Observations reported here were obtained at the MMT Observatory, a joint facility of the University of Arizona and the Smithsonian Institution.  These observations were also based in part on observations made with the NASA Galaxy Evolution Explorer.  GALEX is operated for NASA by the California Institute of Technology under NASA contract NAS5-98034.}
\altaffiltext{2}{George P. and Cynthia Woods Mitchell Institute for Fundamental Physics and Astronomy, Department of Physics and Astronomy, Texas A\&M University, College Station, TX 77843}
\altaffiltext{3}{School of Earth and Space Exploration,  Arizona  State University,  Tempe, AZ  85287} 
\altaffiltext{4}{Center for Astrophysics and Space Sciences, University of California San Diego, La Jolla, CA 92093} 
\altaffiltext{*}{stevenf@physics.tamu.edu}

\begin{abstract}
We present the results of an optical spectroscopic study of 12 {\it GALEX}--discovered star-forming \lya~emitting galaxies (LAEs) at z $\sim$ 0.3.  We measure the emission line fluxes from these galaxies by fitting their observed spectra to stellar population models in order to correct for underlying stellar absorption.  We revisit earlier stellar population model fitting results, finding that excluding now-known AGNs lowers the typical stellar population age and stellar mass of this sample to $\sim$ 300 Myr and $\sim$ 4 $\times$ 10$^{9}$ $\mathcal{M}$\sol, respectively.  We calculate their dust extinction using the Balmer decrement, and find a typical visual attenuation of A$_\mathrm{V}$ $\sim$ 0.3 mag, similar to that seen in many high-redshift LAEs.  Comparing the ratio of \lya/H$\alpha$ and the \lya\ equivalent widths to the measured dust extinction, we find that the ISMs in these objects appear to be neither enhancing nor seriously attenuating the \lya\ equivalent widths, as would be the case in a quasi-clumpy ISM.  Lastly, we perform a detailed analysis of the gas--phase metallicities of these galaxies, and we find that most galaxies in our sample have $Z$ $\lesssim$ 0.4 $Z$\sol.  We find that at a fixed stellar mass, these low-redshift LAE analogs are offset by $\sim$ 0.6 dex lower in metallicity from the general galaxy population at similar redshifts based on the local mass-metallicity relationship.  This implies that galaxies with \lya\ in emission may be systematically more metal poor than star-forming galaxies at the same stellar mass and redshift, similar to preliminary results at $z \sim$ 2.
\end{abstract}

\keywords{galaxies: evolution}

\section{Introduction}
Galaxies selected on the basis of a bright \lya~emission line were originally thought to be indicative of primordial galaxies undergoing their first burst of star--formation \citep{partridge67}, though they have recently been shown to be a complicated group of objects.  Studies utilizing the technique of spectral energy distribution (SED) fitting have shown that typical narrowband--selected \lya~emitting galaxies (LAEs) appear to be predominantly young and low--mass, with ages $<$ 100 Myr and masses $\lesssim$ a few $\times$ 10$^{9}$ $\mathcal{M}$\sol\ \citep[e.g.,][]{gawiser06a, finkelstein07, pirzkal07, lai07, nilsson07a}.  However, in many cases these galaxies also appear to contain some dust, thus they are likely not primordial in nature \citep[e.g.,][]{pirzkal07, lai08, finkelstein08, finkelstein09a, pentericci09}.  In addition, a small fraction of LAEs appear to be more evolved with ages of $\sim$ 0.5 Gyr and masses of $\sim$ 10$^{10}$ $\mathcal{M}$\sol, suggesting that there may be multiple populations of LAEs, or perhaps a tail in the distribution of LAE properties toward more evolved objects \citep[e.g.,][]{finkelstein09a, lai08, pentericci09}.  These more evolved LAEs provide a link to the characteristically more evolved Lyman break galaxies \citep[LBGs; e.g.,][]{kornei10}.

While the stellar masses of galaxies can be reasonably constrained from SED fitting, the remaining properties suffer from degeneracies, and can be poorly constrained in the absence of rest--frame optical detections, especially at high--redshift.  Specifically, the derived ages, dust extinctions and metallicities are only weakly constrained, as they all result in a reddening of the integrated light from a given galaxy.  While photometry spanning the 4000 \AA\ break can improve the fidelity of age estimates, the extinction can typically only be roughly constrained, and the metallicity not constrained at all.  In order to obtain more robust estimates of the physical make--up of LAEs, direct measurements of the extinction and metallicity are necessary.  At low redshift, this is typically done using measurements of the flux of nebular emission lines in the rest--frame optical, e.g., using Balmer line ratios to measure the dust extinction, and ratios of metal lines to measure the gas--phase metallicities \citep[e.g.][]{kobulnicky99, pettini04}.  These analyses are not currently possible at z $\gtrsim$ 3, as these diagnostic lines are shifted into the mid--infrared.  From 2 $<$ z $<$ 3, these measurements are possible using near--infrared spectroscopy on 8--10m class telescopes \citep[e.g.,][]{erb06b, finkelstein10d}.  However, samples of LAEs at these redshifts are only now being compiled, and the required integration times are long.

Presently, we can use low--redshift analogs to try to understand LAEs at high--redshift.  Locally, the \lya\ properties of star-forming galaxies have been studied in great detail.  It was found that \lya\ emission can correlate with bright UV (and H$\alpha$) emission in some regions of a galaxy, but not in others.  It was also shown that \lya\ can escape from galaxies even with a significant dust content, especially if outflows are present in the interstellar medium \citep[ISM; e.g.][]{kunth98, hayes07, atek08, ostlin09}.  These results indicate that the mechanisms which regulate \lya\ escape are complicated, which is an important detail to consider when interpreting the results of high-redshift LAEs.

A little further out, \citet{deharveng08} published the discovery of $\sim$ 100 LAEs at $z \sim$ 0.3 using the space--based {\it Galaxy Evolution Explorer} ({\it GALEX}) telescope.  A study of their luminosity function found that LAEs at low redshift are more rare and less luminous in \lya\ than at $z >$ 3 \citep{deharveng08, cowie10}.  In a followup study, \citet{finkelstein09c} studied the stellar populations of a subsample of 30 of these LAEs, finding that they appear older and more massive than typical high-redshift LAEs.  A higher fraction of low-redshift LAEs appear to host active galactic nuclei (AGN), from $\sim$ 15 -- 40\% \citep{finkelstein09e, scarlata09, cowie10}.  \citet{cowie10} compare the metallicities from the ratio of [N\,{\sc ii}]/H$\alpha$ emission from $z \sim$ 0.3 LAEs to those of $z \sim$ 0.3 continuum-selected galaxies, finding that while the distributions overlap, the LAEs extend to lower metallicties.  Lastly, \citet{atek09} used H$\alpha$ and H$\beta$ observations from a subset of these galaxies to study the \lya\ escape fraction, finding clear evidence for a decrased escape fraction with increased extinction, although some galaxies do exhibit a \lya\ escape fraction greater than the continuum, which could imply a clumpy ISM geometry \citep[e.g.][]{neufeld91, finkelstein09a}.

In this Paper we present the spectra of our sample of 12 {\it GALEX} LAEs in the Extended Groth Strip (EGS) which are dominated by star-formation activity, first analyzed in \citet{finkelstein09c} via SED-fitting, and later shown to be devoid of AGN activity in \citet{finkelstein09e}.  In \S 2 of this Paper, we present the full spectra of every object in our sample, along with their measured line fluxes.  In \S 3, we revisit the typical ages and masses of this sample, comparing to what we earlier derived in \citet{finkelstein09c} when knowledge of AGN activity was unknown.  We discuss the dust properties and insights into the ISM geometries of our sample in \S 4, and in \S 5, we present metallicity measurements with three separate metallicity indicators, as well as examine the mass-metallicity relation, and the implications it has on LAEs near and far.  In \S 6 we present our conclusions.  Where applicable, we assume H$_\mathrm{o}$ = 70 km s$^{-1}$ Mpc$^{-1}$, $\Omega_{m}$ = 0.3 and $\Omega_{\Lambda}$ = 0.7.

\section{Data}
\subsection{Observations and Data Reduction}
We obtained spectroscopy of 23 of the 27 previously analyzed LAEs from \citet{finkelstein09c} in the EGS, using the Hectospec spectrograph with the 6.5m MMT telescope (all 27 could not be observed in one configuration, thus this choice of 23 allowed the most simultaneous observations).  Hectospec is a multi--fiber spectrograph, with 300 1.5\arcs~fibers covering a 1$^\mathrm{o}$ diameter field--of--view at the f/5 focus \citep{fabricant05}.  We acquired spectroscopy for our sample using MMT+Hectospec observations, taken in queue mode on 19 March 2009, obtaining 120 minutes of low--resolution grating spectroscopy, which provides spectral coverage from $\sim$ 3650 -- 9200 \AA, with a spectral resolution of $\sim$ 5 \AA.  The seeing throughout the night was consistently between 0.6\arcs\ -- 0.8\arcs, and thus much smaller than the Hectospec fibers.

There are two primary pipelines for the reduction of Hectospec data, IDL and IRAF\footnote[1]{IRAF is distributed by the National Optical Astronomy Observatory (NOAO), which is operated by the Association of Universities for Research in Astronomy, Inc.\ (AURA) under cooperative agreement with the National Science Foundation.} based.  We chose to use the IRAF--based pipeline, as it has recently been ported for external use as External SPECROAD (or E--SPECROAD)\footnote[2]{E--SPECROAD is available at:\\ http://iparrizar.mnstate.edu/$\sim$juan/research/ESPECROAD/index.php}.  The reduction consists of three tasks: The first, called as {\tt specroad}, makes the master bias, flat and dark frames.  The next task, {\tt specroadcal}, traces the fibers across the chips, creates a dispersion function, and creates the fiber throughput correction.  {\tt Specroadcal} prompts the user to interactively fit a wavelength solution, using lines of helium, neon and argon.  We obtained rms residuals from 0.1 -- 0.15 \AA~using 55 lines spanning the entire wavelength range.  {\tt Specroadcal} then applies a dispersion correction to the flat frame.  The last task, {\tt specroadobj}, reduces all of the object data, using the calibrations made with the previous two tasks, and subtracting the sky which was computed from among the six closest of $\sim$ 50 sky fibers placed throughout the Hectospec field (where the flux variation between the fibers was small).  The software then splits off the 300 fibers into individual fits files, which we used moving forward.  Each of these files contains the wavelength, flux, and flux error array for the specific fiber, where the error spectrum is computed from a variance array that was created from the raw frame and processed through the same reduction steps.

In addition to our primary targets, we also placed fibers on 12 F--type stars from the Sloan Digital Sky Survey, which we used to flux calibrate the data.  These stars are good calibrators, as they are a compromise between strong hydrogen lines in hotter stars and other atomic lines in cooler stars.  For each star, we scaled up a Kurucz (1993) F0V star model to match the fluxes, integrating the model spectrum through the Sloan Digital Sky Survey (SDSS) g$^{\prime}$, r$^{\prime}$ and i$^{\prime}$ filter curves (these three bands were fully encompassed by our wavelength range).  We then computed a scale factor for each filter by dividing the model flux from the SDSS catalog flux for the given star.  A final scale factor was computed as the mean of the scale factor from the three filters.  This scale was then multiplied into the model spectrum.  Finally, a calibration array was computed as the scaled up Kurucz model divided by the observed stellar spectrum.  This was repeated for each star, with a final calibration array coming from the mean of each of the 12 observed stars.  We estimated the accuracy of our flux calibration as the standard deviation of the mean of these 12 observations, which we found to be relatively small, at $<$ 8\%.  The final calibration array was multiplied into our object spectra, which in addition to flux calibration also corrects for the grating transmission function.  The full spectra of all 12 star-forming LAEs are shown in \fig{fig:spectra}.
\begin{deluxetable*}{ccccccccccc}
\tablecaption{Object Properties and Measured Line Fluxes}
\tablewidth{0pt}
\tablehead{
\colhead{Object} & \colhead{RA} & \colhead{Dec} & \colhead{Redshift} & \colhead{Ly$\alpha$} & \colhead{[O\,{\sc ii}]} & \colhead{H$\beta$} & \colhead{[O\,{\sc iii}]} & \colhead{[O\,{\sc iii}]} & \colhead{H$\alpha$} & \colhead{[N\,{\sc ii}]}\\
\colhead{$ $} & \colhead{(J2000)} & \colhead{(J2000)} & \colhead{$ $} & \colhead{$\lambda$1216} & \colhead{$\lambda$3727} & \colhead{$\lambda$4861} & \colhead{$\lambda$4959} & \colhead{$\lambda$5007}& \colhead{$\lambda$6563} & \colhead{$\lambda$6583}\\
}
\startdata
\phantom{E}EGS7&215.17155&53.11394&0.1996&16.1&9.58 $\pm$ 0.12&5.29 $\pm$ 0.08&7.79 $\pm$ 0.04&24.03 $\pm$ 0.11&19.77 $\pm$ 0.20&0.89 $\pm$ 0.11\\
\phantom{E}EGS8&214.43095&52.46829&0.2078&76.7&2.25 $\pm$ 0.08&0.91 $\pm$ 0.06&0.86 $\pm$ 0.02&2.52 $\pm$ 0.07&2.78 $\pm$ 0.10&$<$ 0.10\\
EGS10&214.55878&52.39549&0.2102&12.3&4.80 $\pm$ 0.12&1.74 $\pm$ 0.06&2.01 $\pm$ 0.03&5.93 $\pm$ 0.08&6.02 $\pm$ 0.16&$<$ 0.12\\
EGS11&214.52132&52.75198&0.2440&27.8&2.05 $\pm$ 0.09&0.84 $\pm$ 0.06&0.28 $\pm$ 0.01&1.06 $\pm$ 0.07&2.98 $\pm$ 0.10&0.84 $\pm$ 0.11\\
EGS12&214.70103&52.29891&0.2395&15.2&2.59 $\pm$ 0.08&1.01 $\pm$ 0.05&0.70 $\pm$ 0.02&2.12 $\pm$ 0.06&3.45 $\pm$ 0.12&0.44 $\pm$ 0.07\\
EGS13&214.22613&52.41111&0.2462&35.9&10.52 $\pm$ 0.13&4.59 $\pm$ 0.08&2.73 $\pm$ 0.03&8.43 $\pm$ 0.10&17.95 $\pm$ 0.21&2.59 $\pm$ 0.17\\
EGS19&215.18036&52.71884&0.2466&28.7&3.13 $\pm$ 0.08&1.46 $\pm$ 0.06&0.47 $\pm$ 0.02&1.40 $\pm$ 0.06&5.10 $\pm$ 0.15&1.51 $\pm$ 0.18\\
EGS20&214.81132&52.39066&0.2524&14.2&4.65 $\pm$ 0.09&1.53 $\pm$ 0.07&1.66 $\pm$ 0.03&4.80 $\pm$ 0.08&4.43 $\pm$ 0.11&$<$ 0.21\\
EGS21&215.18599&52.83513&0.2515&25.3&1.39 $\pm$ 0.08&0.77 $\pm$ 0.07&0.08 $\pm$ 0.02&0.25 $\pm$ 0.06&2.62 $\pm$ 0.16&1.23 $\pm$ 0.11\\
EGS22&214.30073&52.59907&0.2607&20.8&1.47 $\pm$ 0.08&0.64 $\pm$ 0.08&0.13 $\pm$ 0.02&0.40 $\pm$ 0.06&2.93 $\pm$ 0.18&$<$ 0.14\\
EGS23&215.35227&52.65558&0.2578&18.0&9.75 $\pm$ 0.11&4.22 $\pm$ 0.09&4.09 $\pm$ 0.04&12.43 $\pm$ 0.10&15.27 $\pm$ 0.20&1.66 $\pm$ 0.14\\
EGS25&214.73180&52.82437&0.2630&20.1&13.89 $\pm$ 0.14&8.84 $\pm$ 0.10&11.81 $\pm$ 0.04&42.91 $\pm$ 0.17&31.02 $\pm$ 0.24&1.66 $\pm$ 0.13\\
\enddata
\tablecomments{All line fluxes are in units of 10$^{-16}$ erg s$^{-1}$ cm$^{-2}$.  When [N\,{\sc ii}] is not detected, we report the 3 $\sigma$ upper limit, as measured from placement of mock lines in the spectra.  The \lya~line fluxes are measured from GALEX spectra in \citet{deharveng08} (which have a characteristic uncertainty of 4 $\times$ 10$^{-16}$ erg s$^{-1}$ cm$^{-2}$).  The redshift is the weighted mean of all lines detected at $\geq$ 3 $\sigma$ significance, with a characteristic error $\Delta$z $\sim$ 0.00001.  The last column is the best-fiitting extinction on the stellar continuum from the spectral fitting described in \S 2.2.}
\end{deluxetable*}

\subsection{Spectral Fitting}
We measure the emission-line fluxes of the objects in our sample using the method described by \citet{moustakas10}.  Briefly, we fit a non-negative linear combination of \citet{bruzual03} population synthesis models with instantaneous burst ages ranging from $5$~Myr to $10.5$~Gyr (the age of the Universe at $z\sim0.25$), assuming the \citet{chabrier03} initial mass function from $0.1-100~M_{\odot}$.  Motivated by the fact that the gas-phase metallicities of these objects are sub-solar (see \S 5), we assume a fixed stellar metallicity of $Z=0.004$ ($20\%Z_{\odot}$); however, we have verified that using solar-metallicity templates has a negligible effect on the derived line-strengths.  After subtracting the best-fitting stellar continuum from the data, we are left with a pure emission-line spectrum self-consistently corrected for underlying stellar absorption.  Finally, we fit all the emission lines simultaneously using Gaussian line-profiles.  In \fig{fig:linefit} we illustrate our fitting method for one object in our sample, EGS19.  In the top panel we plot the observed spectrum (light grey), the best-fitting continuum model (black), and the residual emission-line spectrum (dark grey).  We show the full best-fitting emission-line model spectrum (dark red) in the top panel, and in the lower panel we zoom into the emission lines of interest.  The lower, middle panel in particular emphasizes the importance of correcting the H$\beta$ line-strength for underlying stellar absorption using this kind of technique \citep[see, e.g.,][for a recent review]{walcher10}.

\begin{figure*}
\epsscale{0.95}
\plotone{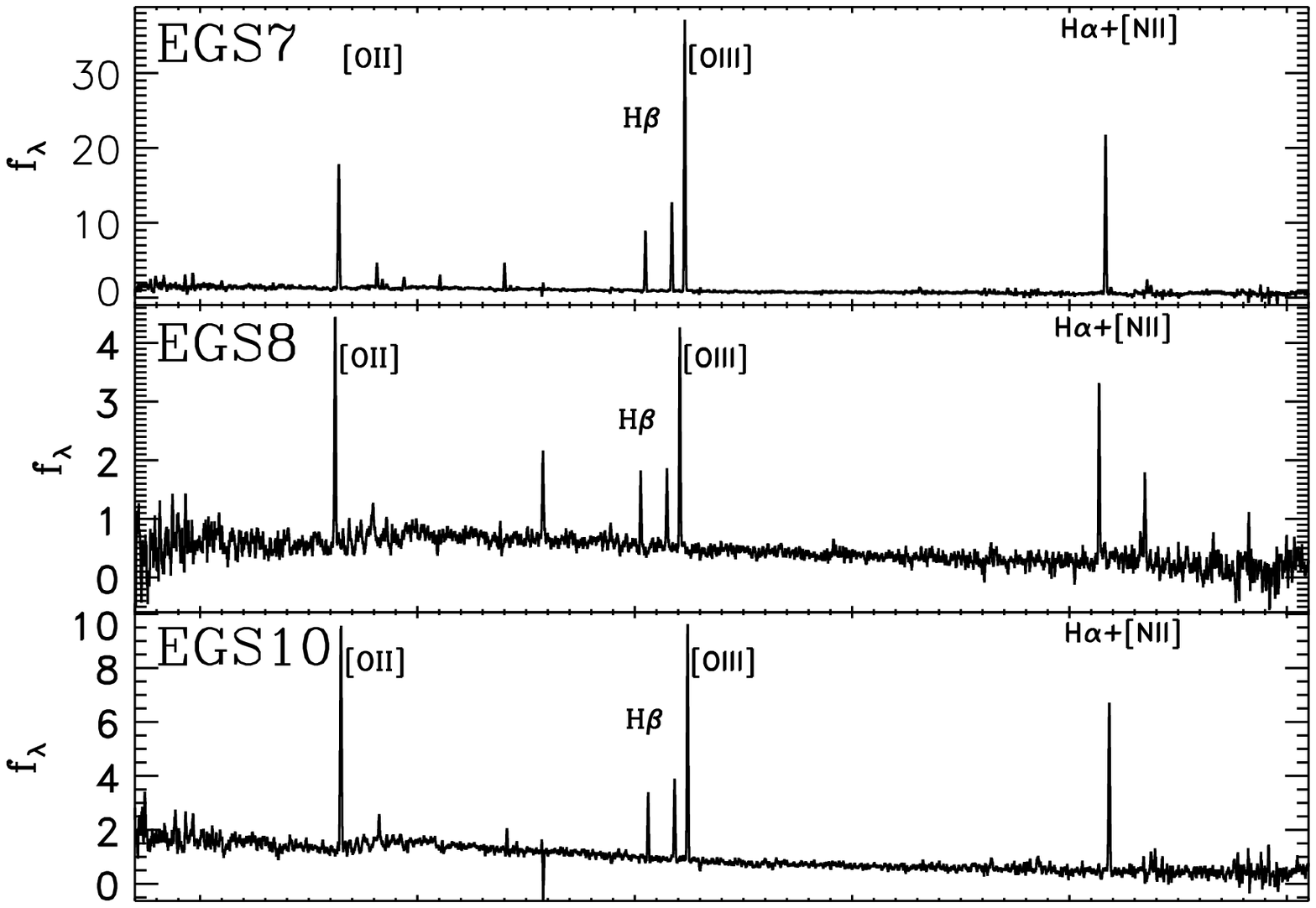}
\vspace{-2.8mm}
\plotone{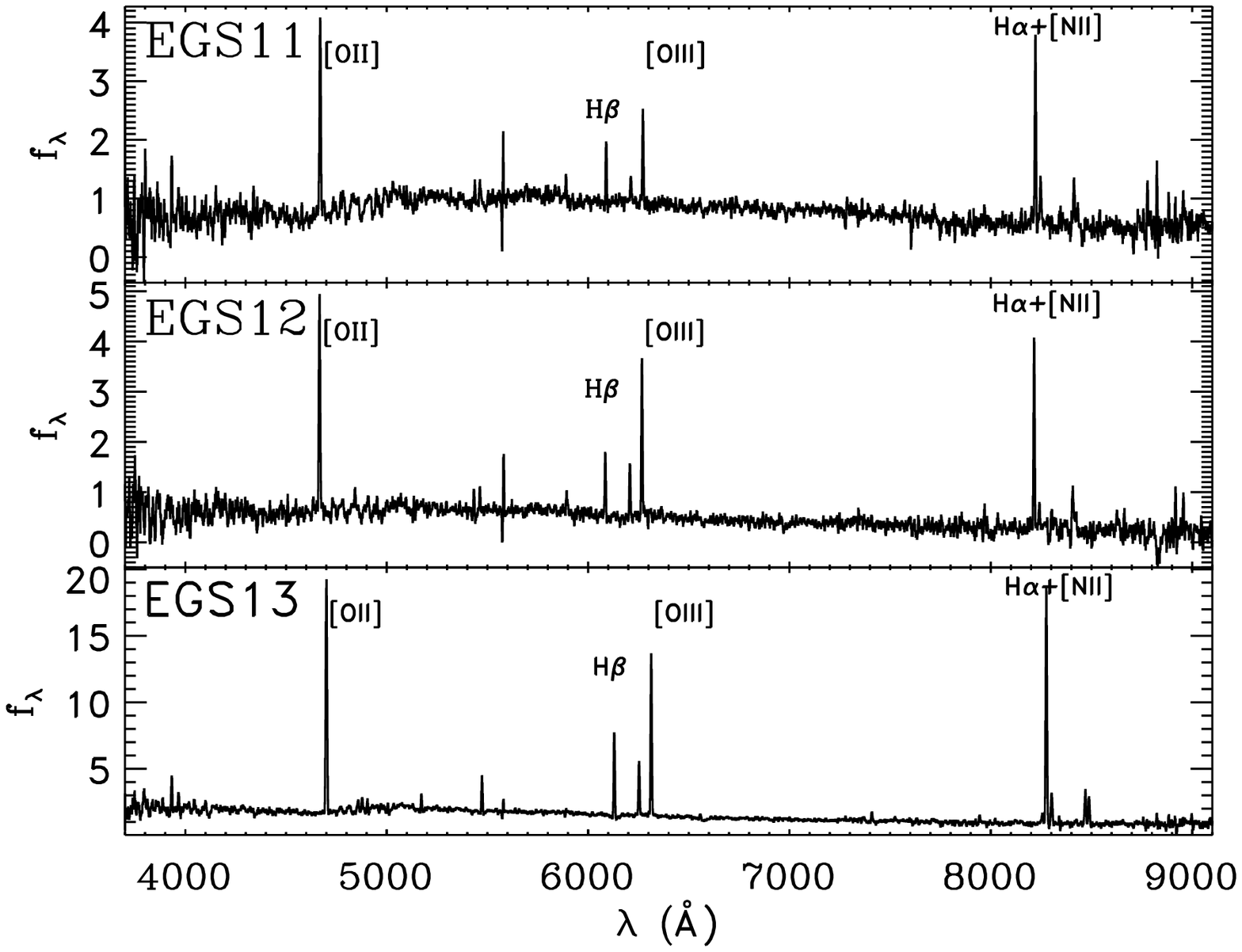}
\vspace{8mm}
\caption{The full Hectospec spectra of the 12 star-forming LAEs in our sample.  The vertical axis is in units of 10$^{-17}$ erg s$^{-1}$ cm$^{-2}$ \AA$^{-1}$, and the reader should note that the vertical scale varies from spectrum to spectrum.  The typical observed line width is $\sim$ 6 \AA.  In some cases the 5577 \AA~atmospheric line is poorly subtracted, and should be ignored.}\label{fig:spectra}
\end{figure*}
\addtocounter{figure}{-1}

\begin{figure*}
\epsscale{0.95}
\plotone{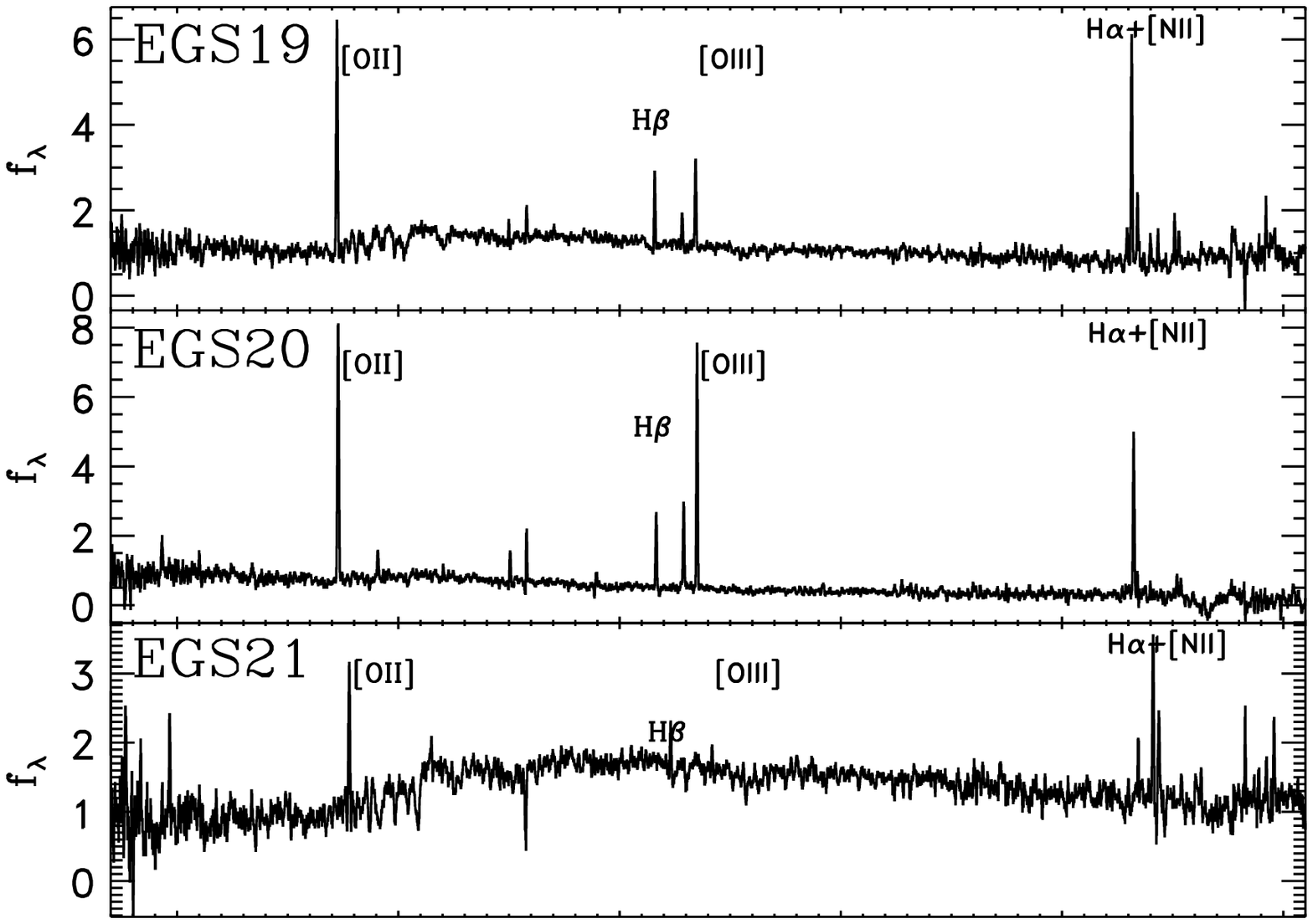}
\vspace{-2.8mm}
\plotone{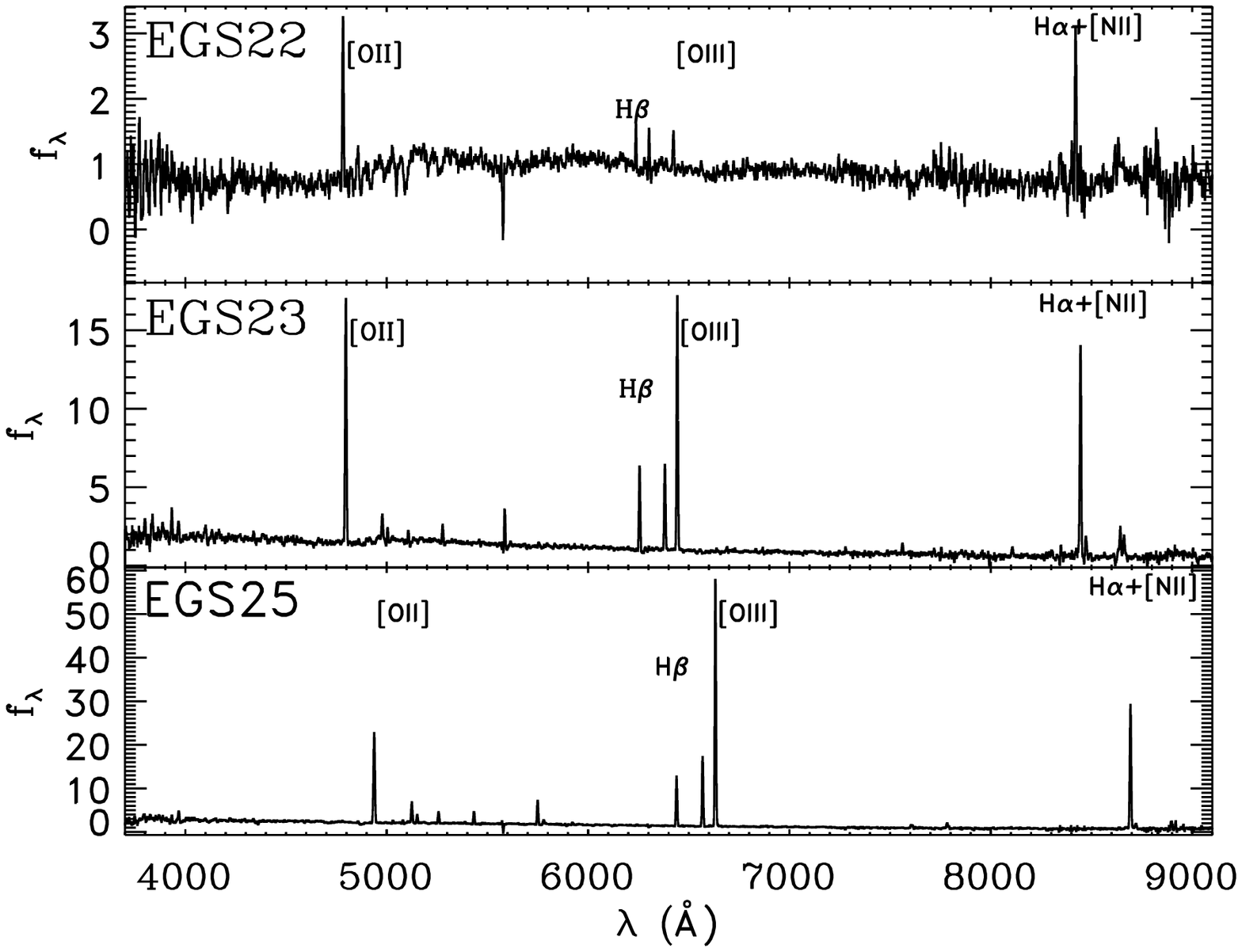}
\vspace{5mm}
\caption{(cont'd)  The remaining spectra in our sample}\label{fig:spectra2}
\end{figure*}

The [N\,{\sc ii}] emission line was not detected in four of our objects.  In these cases, we computed upper limits to their line fluxes by adding in mock emission lines to the spectrum, and repeatedly decreasing their line strength until the signal--to--noise dropped below 3 $\sigma$.  We found that on average, the 3 $\sigma$ line flux limits were $\approx$ 1.0 $\times$ 10$^{-17}$ erg s$^{-1}$ cm$^{-2}$.  In the analysis below, we use the 3 $\sigma$ upper limits as the line fluxes for these lines.  Additionally, the [O\,{\sc iii}] $\lambda$4959 line was not detected in two objects (EGS21 and EGS22).  This is understandable as these two objects had the lowest [O\,{\sc iii}] $\lambda$5007 fluxes, though both were detected at $>$ 4$\sigma$.  We thus set the [O\,{\sc iii}] $\lambda$4959 flux to be 0.336 $\times$ the [O\,{\sc iii}] $\lambda$5007 flux (the theoretical line flux ratio from \citet{storey00}) for these two objects.  We note that out of the remaining 10 objects, all but two have their ratio of [O\,{\sc iii}] flux ratios at $<$ 3$\sigma$ of the theoretical value.  The line fluxes for the measured lines which are relevant to our analysis are given in Table 1.

\section{The Ages and Masses of Star-Forming Low-Redshift LAEs}
In \citet{finkelstein09c} we derived the stellar population properties of the parent sample of 30 LAEs via spectral energy distribution (SED) fitting, from which we drew the 23 objects observed with Hectospec, including the 12 confirmed star-formation-dominated galaxies presented here.  As this was prior to the spectroscopic observations presented in this Paper (and \citet{finkelstein09e}), no knowledge of AGNs or line emission was accounted for in the fitting process.  The model templates used in this procedure account only for stellar emission, thus both AGN continuum emission and nebular line emission can cause incorrect fits.  Although in this paper we exclude the known AGNs from our analysis, it is still prudent to revisit the SED-fitting results as the stellar masses are key to our analysis below.

We have re-performed the SED fitting using the same procedure outlined in \citet{finkelstein09c}.  However, here we use the updated 2007 version of the models of \citet{bruzual03}.  In brief, we generate a suite of metals, varying the metallicity (0.02 $\leq Z \leq$ 2.5 $Z$\sol), star-formation history (exponentially decaying with 10$^{5}$ $\leq \tau_{SFH} \leq$ 10$^{9.6}$), stellar population age (10 Myr -- t$_{universe}$[z]), dust extinction (0 $\leq$ A$_\mathrm{V}$ $\leq$ 1.6 mag) and the ISM clumpiness parameter q (0 $\leq$ q $\leq$ 10); where q = $\tau_{Ly\alpha}$/$\tau_{continuum}$.  We fit this suite of models to our observations; GALEX far-ultraviolet and near-ultraviolet (FUV and NUV, respectively), CFHT Legacy Survey u$^{\prime}$, g$^{\prime}$, r$^{\prime}$, i$^{\prime}$ and z$^{\prime}$, and the GALEX \lya\ line flux.  We find the best-fit model by comparing flux ratios of all bands to the u$^{\prime}$-band (i.e. colors), and the calculate the stellar mass of the best-fit model by a weighted mean of the observed fluxes to the fluxes of the best-fit model.  Uncertainties on the best-fit parameters (excepting stellar mass) were computed via a Bayesian likelihood analysis \citep{kauffmann03a}, where the probability distribution is related to the full $\chi^2$ array via P $\propto$ e$^{-\chi^{2}/2}$.  The $\chi^{2}$ array is five-dimensional -- one for each of our free parameters (5 metallicities $\times$ 5 SFHs $\times$ 124 ages $\times$ 21 values of extinction $\times$ 10 values of q).  Using age as an example, the probability of a given object having a given age is simply the (normalized) total of P over all other free parameters.  As the mass is determined after the selection of the best-fit model, we determine the uncertainty on the masses via 10$^{2}$ Monte Carlo simulations.  The emission-line subtracted object magnitudes, best-fit stellar population ages, stellar masses and dust extinction, as well as the 68\% confidence ranges on these parameters are given in Table 2.

\begin{figure}
\epsscale{1.15}
\vspace{2mm}
\plotone{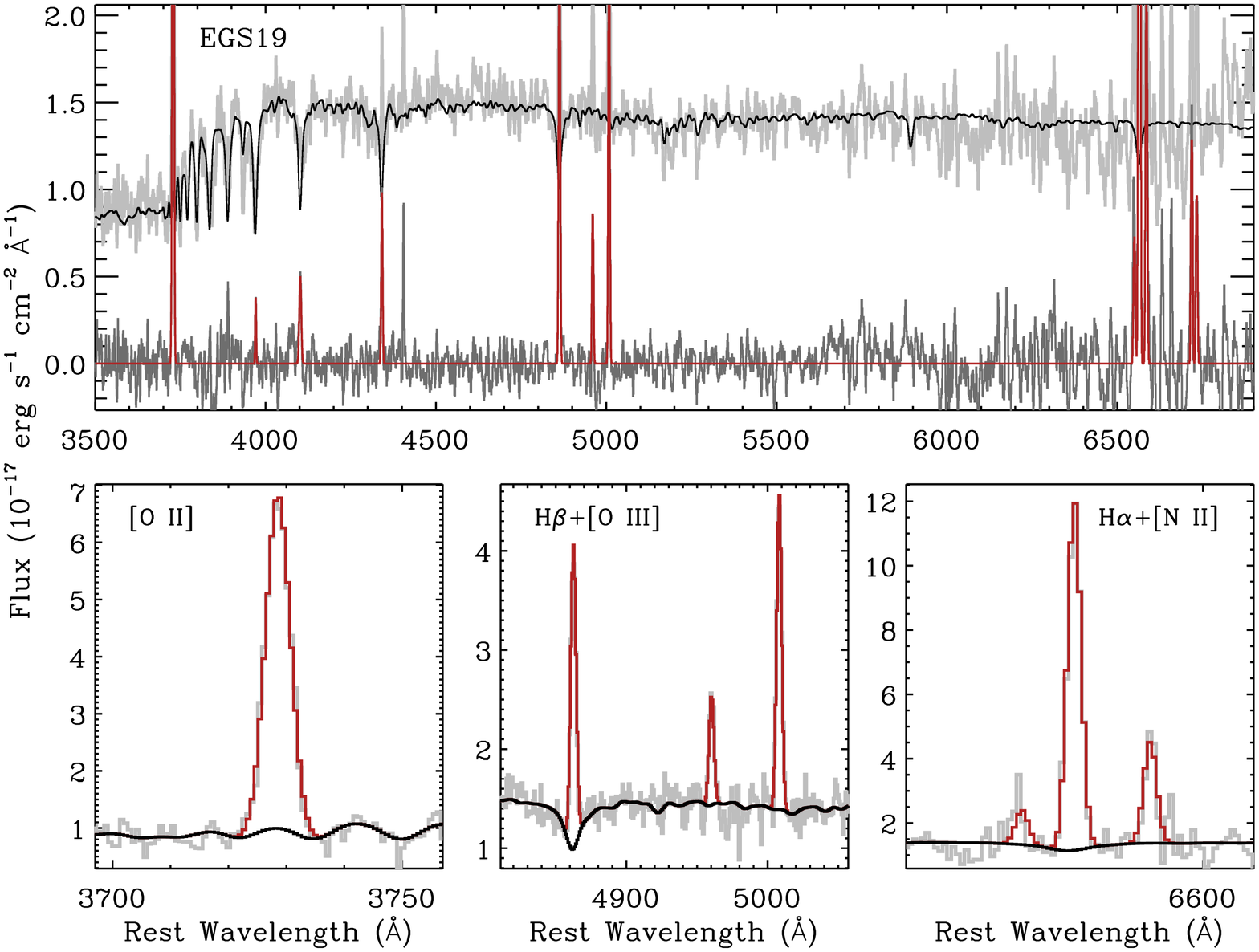}
\vspace{2mm}
\caption{An example of the emission-line fitting method on EGS19.  The top panel shows our observed spectrum in gray (smoothed by a 5-pixel boxcar), with the best-fitting spectral model shown as the black curve.  The underlying emission line spectrum, after subtracting the stellar continuum, is shown in red.  The bottom three panels show the regions around some of the lines of interest to our study.  Of note  is the strong stellar Balmer-line absorption.  Correcting for this underlying absorption is crucial to an accurate measurement of the Balmer decrement.}\label{fig:linefit}
\end{figure}
 
\begin{figure*}[!ht]
\epsscale{0.5}
\plotone{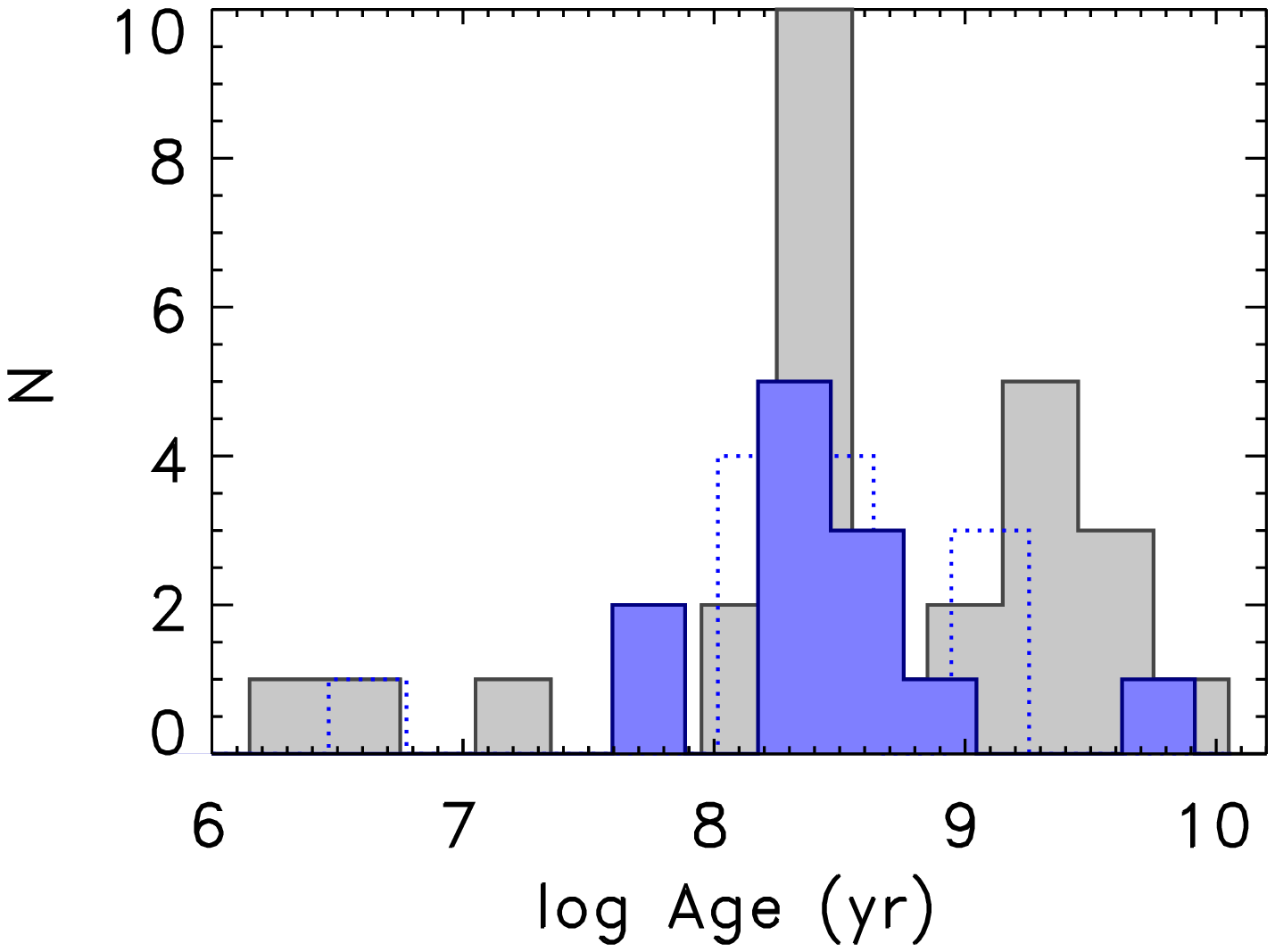}
\plotone{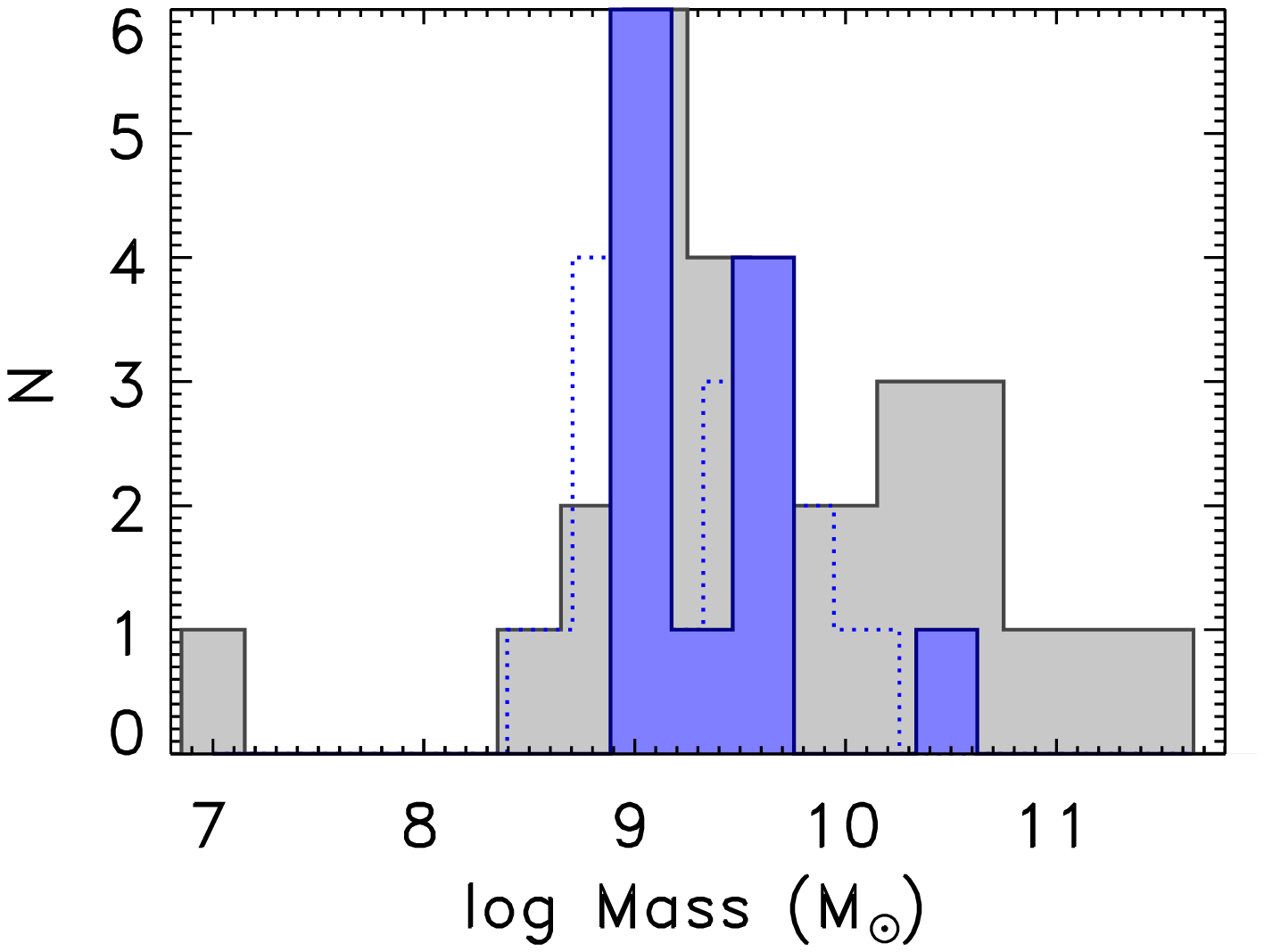}
\vspace{5mm}
\vspace{-1mm}
\caption{Left: Distribution of best-fit stellar populations ages for the 27 LAEs from \citet{finkelstein09c} are shown as the gray histogram.  The best-fit ages we derive here for the subsample of 12 star-forming LAEs are shown as the blue histogram.  While the original sample of 27 LAEs had many objects with best-fit ages $>$ 1 Gyr, when we exclude now-known AGNs from \citet{finkelstein09e}, we find that all but one galaxy appears younger than 1 Gyr.  The dotted blue histogram denotes the ages for the same 12 LAEs from \citet{finkelstein09c}, which broadly agree with our updated results.  Right: Same as the left, for the derived stellar masses.  Similar to the ages, a smaller characteristic mass appears likely for these low-redshift LAEs once known AGNs are excluded.}\label{fig:dists}
\end{figure*}

Histograms comparing our best-fit stellar population ages and stellar masses are shown in \fig{fig:dists}, compared to the results for the full sample from \citet{finkelstein09c}.  While our earlier study found a number of objects with ages in excess of a Gyr, those primarily appear to be dominated by AGN-hosting LAEs, thus the ages may be unreliable.  We find that the updated ages for our star-forming LAE sample are primarily between 100 -- 1000 Myr, with a median of $\sim$ 300 Myr.  We find a similar result for the stellar masses, shown in the right-hand plot of \fig{fig:dists}, in that the many LAEs with derived stellar masses $>$ 10$^{10}$ $\mathcal{M}$\sol\ appear to be dominated by AGN hosts, and that the star-forming LAEs in our current sample have masses $\sim$ 10$^{9}$ -- 10$^{10}$ $\mathcal{M}$\sol\ (with a median of 4 $\times$ 10$^{9}$ $\mathcal{M}$\sol).  While these low-redshift LAEs are older and more massive than typical narrowband selected LAEs, they are similar to those of the more evolved, rarer, LAEs seen at high redshift \citep[e.g.,][]{finkelstein09a, lai08, pentericci09}.  However, we do caution that the derived ages are highly uncertain, as although many objects have best-fit ages on the order of a few hundred Myr, the uncertainties on these derived ages are high - sometimes orders of magnitude.  As shown in Table 2, the uncertainties on the stellar masses are much smaller, and we use these best-fit masses with their uncertainties in our metallicity analysis in \S 5.

\begin{deluxetable*}{ccccccccccccccc}
\tabletypesize{\footnotesize}
\tablecaption{SED Fitting Results}\label{tab:appendix}
\tablewidth{0pt}
\tablehead{
\colhead{Object} & \colhead{m$_{FUV}$} & \colhead{m$_{NUV}$} & \colhead{m$_{u}$} & \colhead{m$_{g}$} & \colhead{m$_{r}$} & \colhead{m$_{i}$} & \colhead{m$_{z}$} & \colhead{f$_{Ly\alpha}$/10$^{-15}$} & \colhead{Age} & \colhead{Age} & \colhead{Mass} & \colhead{Mass} & \colhead{A$_\mathrm{V}$} & \colhead{A$_\mathrm{V}$}\\
\colhead{$ $} & \colhead{(mag)} & \colhead{(mag)} & \colhead{(mag)} & \colhead{(mag)} & \colhead{(mag)} & \colhead{(mag)} & \colhead{(mag)} & \colhead{(erg s$^{-1}$ cm$^{-2}$)} & \colhead{Best-Fit} & \colhead{68\% Range} & \colhead{Best-Fit} & \colhead{68\% Range} & \colhead{Best-Fit} & \colhead{68\% Range}\\
\colhead{$ $} & \colhead{$ $} & \colhead{$ $} & \colhead{$ $} & \colhead{$ $} & \colhead{$ $} & \colhead{$ $} & \colhead{$ $} & \colhead{$ $} & \colhead{(Myr)} & \colhead{(Myr)} & \colhead{(10$^{9}$ $\mathcal{M}$\sol)} & \colhead{(10$^{9}$ $\mathcal{M}$\sol)} & \colhead{(mag)} & \colhead{(mag)}\\
}
\startdata
EGS7&21.63&21.29&21.31&20.88&20.57&20.45&20.48&2.8&450&\phantom{0}260 -- \phantom{0}800&\phantom{0}1.3&\phantom{0}1.2 -- \phantom{0}1.5&0.32&0.08 -- 0.57\\
EGS8&22.03&21.66&21.88&21.18&20.76&20.64&20.63&1.5&320&\phantom{0}140 -- \phantom{0}400&\phantom{0}1.6&\phantom{0}1.6 -- \phantom{0}2.1&0.00&0.00 -- 0.40\\
EGS10&21.59&21.29&21.45&21.02&20.75&20.64&20.60&2.9&570&\phantom{0}320 -- \phantom{0}900&\phantom{0}1.6&\phantom{0}1.3 -- \phantom{0}1.7&0.32&0.00 -- 0.49\\
EGS11&22.16&21.77&21.34&20.58&20.07&19.85&19.73&1.4&290&\phantom{00}80 -- \phantom{0}360&\phantom{0}3.8&\phantom{0}3.7 -- \phantom{0}4.9&0.49&0.32 -- 1.05\\
EGS12&21.64&21.35&21.14&20.42&19.93&19.76&19.64&2.5&320&\phantom{00}90 -- 2600&\phantom{0}4.9&\phantom{0}4.8 -- \phantom{0}5.8&0.49&0.16 -- 0.89\\
EGS13&21.76&21.23&21.24&20.61&20.13&19.93&19.80&2.1&810&\phantom{0}400 -- 2600&\phantom{0}4.2&\phantom{0}3.8 -- \phantom{0}7.3&0.57&0.16 -- 0.81\\
EGS19&21.86&21.29&21.28&20.53&20.03&19.82&19.67&2.2&320&\phantom{0}180 -- 4200&\phantom{0}4.9&\phantom{0}4.8 -- 10.8&0.57&0.16 -- 0.89\\
EGS20&22.02&21.64&21.71&21.21&20.81&20.74&20.66&1.8&720&\phantom{0}230 -- 1600&\phantom{0}2.0&\phantom{0}1.7 -- \phantom{0}2.0&0.32&0.08 -- 0.49\\
EGS21&21.75&21.28&21.11&20.30&19.58&19.29&19.11&1.9&9000&2600 -- 8800&34.4&10.5 -- 33.7&0.00&0.00 -- 0.65\\
EGS22&21.99&21.49&21.32&20.62&20.02&19.82&19.67&1.7&260&\phantom{0}160 -- 3500&\phantom{0}4.8&\phantom{0}4.7 -- \phantom{0}4.9&0.65&0.24 -- 0.89\\
EGS23&21.88&21.37&21.50&21.15&20.82&20.63&20.55&2.9&80&\phantom{00}27 -- \phantom{0}570&\phantom{0}1.1&\phantom{0}1.1 -- \phantom{0}2.0&0.97&0.32 -- 1.38\\
EGS25&22.18&21.10&21.44&21.17&20.95&20.87&20.65&2.6&60&\phantom{00}18 -- \phantom{00}71&\phantom{0}1.2&\phantom{0}1.2 -- \phantom{0}2.2&0.97&0.89 -- 1.46\\
\enddata
\tablecomments{These magnitudes have had all emission lines detected at $\geq$ 3$\sigma$ subtracted.  The magnitude errors are 0.2 and 0.02 mag in the GALEX and optical bands, respectively (see \citet{finkelstein09c}).}
\end{deluxetable*}

\section{Dust Extinction and the ISM}
Using the Balmer decrement we can measure the level of dust extinction in our galaxies, providing a much more robust estimate than that derived from SED fitting (listed in Table 2), which is subject to a number of degeneracies.  Traditionally, this is done by measuring the ratio of H$\alpha$/H$\beta$, and then comparing to the theoretical ratio, which we take as H$\alpha$/H$\beta =$ 2.86 \citep{osterbrock89}.  Any increase in this ratio over the theoretical value is likely due to dust, and, assuming an extinction law, one can compute the color excess.  Table 3 tabulates the extinction results assuming the \citet{calzetti00} starburst extinction law.  The computed color excesses for our objects range from 0 $\lesssim$ E(B-V)$_{g}$ $\lesssim$ 0.4, with 10/12 objects having E(B-V)$_{g}$ significantly greater than zero (at 1$\sigma$ significance; 9/12 at 2$\sigma$).  We converted the color excess on the nebular emission to an extinction on the stellar continuum assuming that E(B-V)$_{stellar}$ = 0.44 E(B-V)$_{gas}$, as has been found locally \citep{calzetti00}.  We then converted to the stellar attenuation in the V-band via the adopted attenuation curve.  The derived attenuation on the stellar continuum ranges from A$_\mathrm{V}$ = 0 -- 0.7, with a mean A$_\mathrm{V}$ $\sim$ 0.3.  Comparing the derived attenuations from the Balmer decrement to those from the photometric SED fitting (listed in Table 2), we find excellent agreement, as the difference in attenuation for 9/12 galaxies are smaller than their combined 1$\sigma$ uncertainties.  This implies that in this sample of galaxies the conversion used from gas extinction to stellar extinction is accurate, and thus that the nebular emission lines are more extincted than the stellar continuum.  However, the uncertainties on the Balmer-decrement-derived attenuations are much smaller than those from SED fitting, highlighting the advantage of spectroscopy.

\citet{atek09} studied the \lya\ escape fraction in LAEs from the z $\sim$ 0.3 sample of \citet{deharveng08}, obtaining spectroscopic measurements of objects from the Chandra Deep Field - South (CDFS) and ELAIS-S1 fields.  As part of their analysis, they also computed the dust extinction using the ratio of H$\alpha$/H$\beta$.  From their Figure 2, they found values of the color excess from $-$0.3 $<$ E(B-V) $<$ 0.5, covering roughly the same range of color excess values as we find in our independent sample.  \citet{atek09} find that 12/21 of their objects are consistent with E(B-V) $>$ 0.  \citet{scarlata09} also studied z $\sim$ 0.3 LAEs, obtaining optical spectroscopy of 31 galaxies drawn from the \citet{deharveng08} EGS, NGPDWS and SIRTFFL samples, with their EGS sample including four objects in our sample of 12 star-forming galaxies.  They do not compute the dust extinction, but to facilitate comparison we have computed the color excess based on their tabulated H$\alpha$ and H$\beta$ fluxes for the 20 objects which have 3 $\sigma$ detections of both emission lines.  Their objects have a range of $-$0.05 $<$ E(B-V) $<$ 0.74, with 11/20 objects consistent with E(B-V) $>$ 0 at $>$ 1 $\sigma$.  While both \citet{atek09} and \citet{scarlata09} find a slightly smaller fraction of objects with significant dust attenuation than we do, this difference is likely due to the small number of objects analyzed in all three studies, as well as the large uncertainties on the extinction in general.  We note that \citet{atek09} used the extinction curve of \citet{cardelli89}.  Had we adopted this curve, we would have found higher values of the color excess, though the derived attenuation would be similar to what is reported in Table 2.

While many high--redshift LAEs typically have derived non--zero extinctions, these are inferred from SED fitting.  As discussed in \citet{finkelstein09c} and \S 3, the galaxies in our sample are more evolved than typical narrowband selected LAEs at high redshift.  Thus, it is interesting to compare the mean value of A$_\mathrm{V}$ $\sim$ 0.3 mag measured here to high--redshift LAEs.  While \citet{gawiser06a} derived no dust extinction in LAEs at z $\sim$ 3.1, others have frequently found evidence for dust in typical narrowband--selected LAEs.  \citet{pirzkal07} found A$_\mathrm{V}$ = 0 -- 0.6 mag in LAEs at z $\sim$ 5, \citet{lai07} found A$_\mathrm{V}$ = 0.4 -- 1.7 mag in LAEs at z $\sim$ 5.7, and \citet{finkelstein09a} found A$_\mathrm{V}$ = 0.1 -- 1.1 mag in LAEs at z $\sim$ 4.5.  Thus, though it appears that although high--redshift LAEs are on average younger and lower mass, many of them appear to contain a similar range of dust attenuation as these low--redshift LAEs.

\subsection{ISM Geometry}
The geometry of the ISM can play a strong role in the escape of \lya~photons.  \lya\ photons are resonantly scattered by H\,{\sc i}, thus in a homogeneous geometry (i.e. a dust screen in front of the sources), where H\,{\sc i} gas and dust are evenly mixed, \lya~has a difficult time escaping, and thus has a much greater chance of encountering a dust grain than the continuum photons.  On the other hand, if the ISM is inhomogeneous, with the dust locked together with neutral gas in many clumps residing in a tenuous, ionized medium, the \lya~photons will resonantly scatter at the edge of these clumps and avoid the dust, while the continuum light will pass through and be subject to dust attenuation \citep{neufeld91, hansen06}.  This has the net effect of increasing the \lya~escape fraction over that of the continuum, and thus increasing the \lya~equivalent width (EW).  This has been used to explain the large number of high \lya~EW LAEs which have been found \citep[e.g.,][]{kudritzki00, malhotra02}.  In \citet{finkelstein08} and \citet{finkelstein09a}, we included this scenario in our stellar population models.  We parameterized the ISM geometry with a free parameter ``$q$'', which is defined by
\begin{equation}
f^{\prime}_{Ly\alpha} = f_{Ly\alpha} \times \mathrm{exp}(-q\tau_c) 
\end{equation}
where $\tau_c$ is the optical depth of dust on the continuum light, thus $q$ = $\tau_{Ly\alpha} / \tau_{c}$.  We make the assumption that the covering fraction of clumps is large, meaning that a \lya\ photon will scatter many times prior to escape.  In this scenario, the extinction curve (which we assume to be that of \citet{calzetti00}) is the same as that for a uniform dust screen \citep{scarlata09}, excepting that \lya\ photons can preferentially avoid encountering the dust.  When $q = 10$, this has the effect of a homogeneous ISM, where \lya~is attenuated far more than the continuum as resonant scattering increases the chance of absorption by dust.  When $q = 0$, this simulates the \citet{neufeld91} scenario, where all of the dust is locked away inside neutral gas clumps.  In between values of $q$ reflect intermediate geometries, with any value of $q < 1$ enhancing the \lya~EW.  Using this, we found that dust enhancement can explain the SEDs of many z $\sim$ 4.5 LAEs, as 10/14 objects had $q < 1$.  We also used the ISM clumpiness parameter $q$ when we fit models in \citet{finkelstein09c} and \S 3 to the present sample.  Our 12 low-redshift LAEs have a best--fit range from 0 $\leq$ $q$ $\leq$ 1, with 5/12 objects having $q$ from 0.75 -- 1.0 (using the updated model fitting results discussed in \S 3), reflecting an ISM which does not much enhance or attenuate the \lya~EWs. 

\begin{figure*}[!ht]
\epsscale{0.5}
\plotone{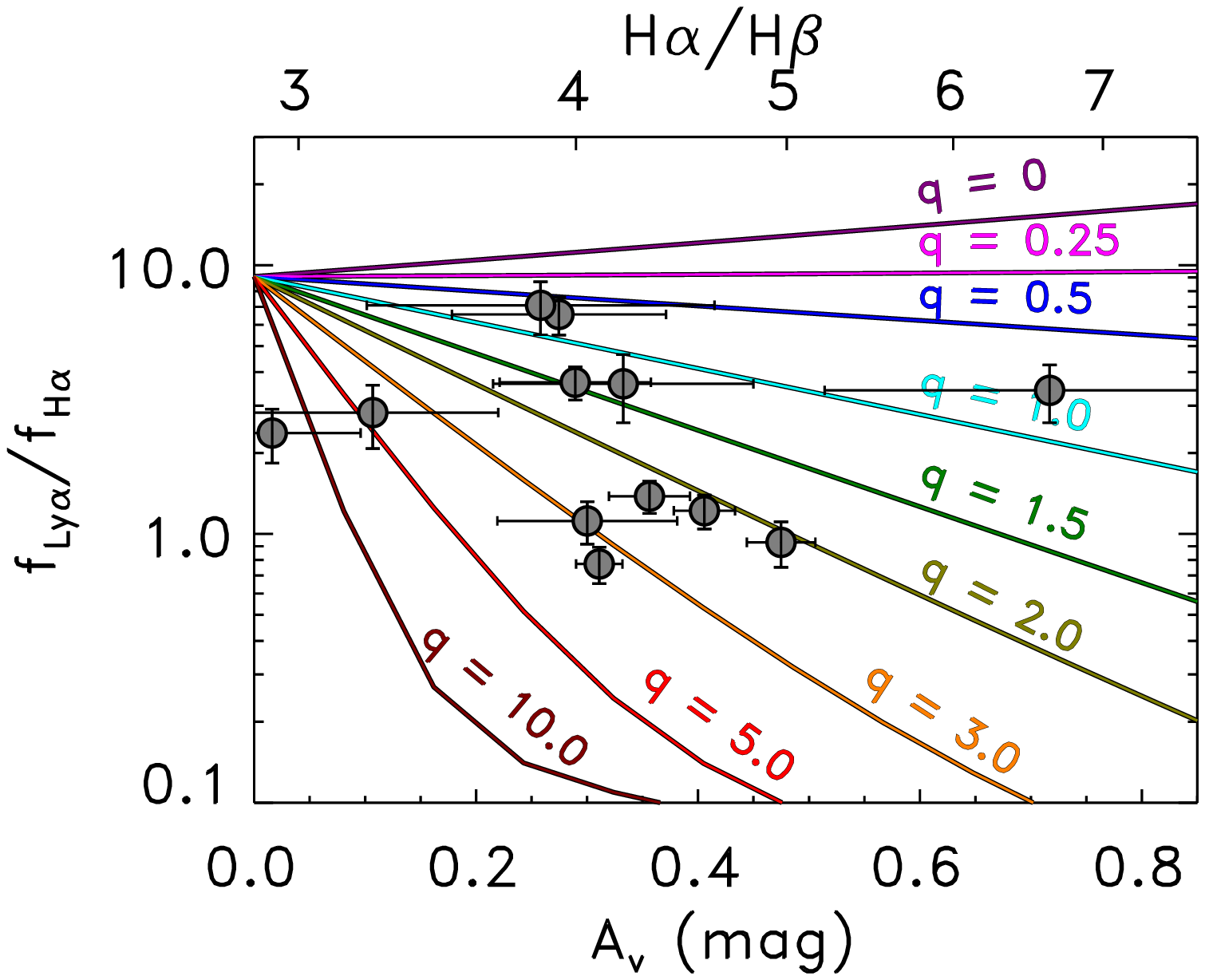}
\plotone{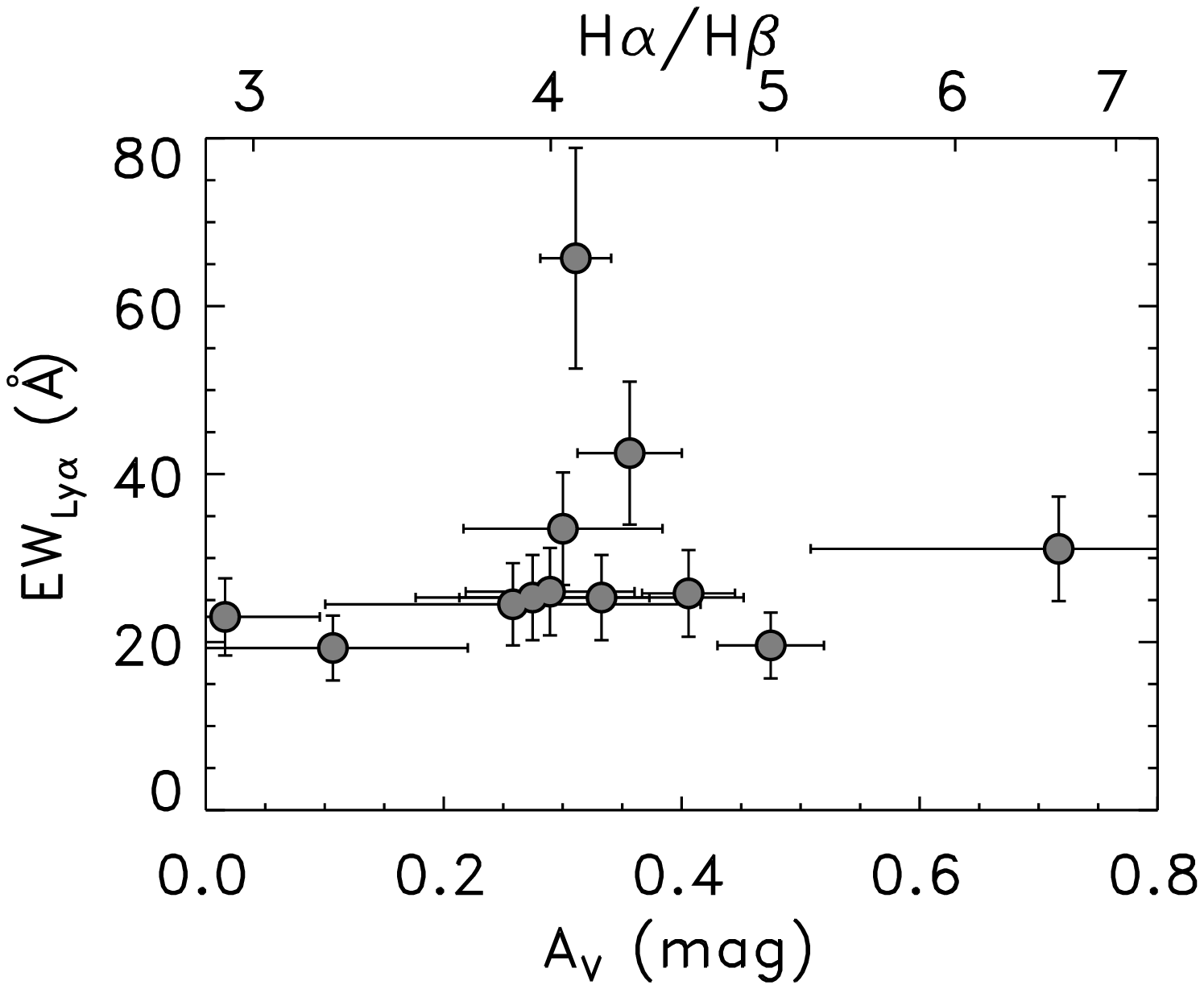}
\vspace{-2mm}
\caption{Left: The distribution of \lya/H$\alpha$ ratios versus the measured dust attenuation for the 12 star--forming LAEs in our sample.  The values of \lya/H$\alpha$ have been corrected for aperture differences.  The colored lines show model tracks for constant values of $q$, the ISM clumpiness factor.  By studying our objects in this plane, we see that few appear to have an ISM geometry that preferentially favors \lya~escape over continuum photons.  Most of the sample lies at q $\sim$ 1--3, implying that \lya\ encounters a similar level of dust (and possibly a bit more) than than the continuum.  Right: The \lya\ EW versus the derived dust attenuation.  The EW is roughly invariant to the level of dust, implying that the ISMs of these objects are neither strongly enhancing nor depleting the \lya\ EWs, and thus that neither a uniform dust screen, or a purely clumpy ISM geometry apply to these objects.
}\label{fig:lyaha}
\end{figure*}

With these spectroscopic data, we can now perform a more in depth analysis into the ISM geometry by comparing the resonantly scattered \lya~line to a non--resonant line, such as H$\alpha$, as a proxy to trace the continuum photons.  Under Case B recombination, the intrinsic ratio of \lya/H$\alpha$ is $\sim$ 8.7.  Any deviation from this value is presumably due to dust (although preferential scattering of \lya\ photons could play a role).  In a typical homogeneous ISM, this value will drop quickly if dust is present.  If, however, the ISM is in a configuration that results in the enhancement of the \lya~EW, this value will increase.  

Our H$\alpha$ measurements come from a 1.5\arcs~diameter fiber, while the Ly$\alpha$ measurements are from the far-ultraviolet (FUV) {\it GALEX} prism, which has a beam full--width at half--maximum (FWHM) of 4.2\arcs~\citep{morrissey07}.  As many of these galaxies are larger than 1.5\arcs~(see Figure 1 of \citet{finkelstein09c}), we need to account for aperture differences between the Ly$\alpha$ and H$\alpha$ measurements.  We did this by first measuring photometry of our sample of galaxies in the Canada--France--Hawaii--Telescope Legacy Survey (CFHTLS) i$^{\prime}$ image (which has a seeing of $\sim$ 0.7\arcs, comparable to our Hectospec observations), using an aperture diameter of the same size as the Hectospec fibers (1.5\arcs).  We then smoothed this image such that the FWHM of stars in the image matched the FWHM of the {\it GALEX} beam, and performed photometry on our sample again, this time using 4.2\arcs~diameter apertures.  In both cases, the photometry was done using the Source Extractor software package \citep{bertin96}.  The aperture correction which we applied to the H$\alpha$ flux measurement is thus the ratio of the flux in the 4.2\arcs~aperture to that in the 1.5\arcs\ aperture.  This aperture correction factor ranged from 1 -- 4, with a median of 1.3, and 11/12 galaxies having an aperture correction factor of $<$ 2.0 (one object had a correction of $<$ 1, which we rounded up to 1).

The top panel in \fig{fig:lyaha} plots the \lya/H$\alpha$ ratios, using the aperture correction, for our sample versus their derived dust attenuation from the previous section.  Examining first the \lya/H$\alpha$ ratios, we find they span 1 $<$ \lya/H$\alpha$ $<$ 7, with all 12 objects having line ratios significantly less than 8.7, implying dust is likely present.  However, the exact amount of dust cannot be derived from this ratio alone as the geometry plays a role.  Using stellar population models from \citet{finkelstein09c}, we also plot tracks of constant ISM clumpiness, from $q$ = 0 -- 10.  Thus, by examining where our objects fall in this plane, we can gain insight into their ISM geometry.  

Investigating this plane, we find that the majority of our objects are consistent with values of q $\sim$ 1--3.  This implies that the dust that is present lies in a geometry somewhere between that of a uniform screen and that of purely clumps.  In the bottom panel of \fig{fig:lyaha}, we plot the {\it GALEX} derived \lya\ EW versus the attenuation we derive here.  We see that the \lya\ is between 20 -- 40 \AA\ for most objects regardless of the level of extinction.  This again implies that the ISM is neither overtly enhancing nor depleting the \lya\ EW (similar also to the results from SED fitting).  Similar results are seen at $z \sim$ 2, where \citet{blanc10} recently analyzed a sample of $\sim$ 100 LAEs and found that \lya\ and the rest-frame UV continuum experienced similar levels of extinction (i.e. q = 1).  As mentioned above, \citet{finkelstein09a} found than many z $\sim$ 4.5 LAEs were consistent with q $\leq$ 1, thus at much higher redshift the ISMs may be more clumpy, though this was indirectly inferred via SED fitting.

\citet{atek09} studied the escape fraction of \lya\ photons in their sample of $z \sim$ 0.3 LAEs, finding that some objects indicate \lya-assisted escape due to a clumpy ISM (q $<$ 0), some objects appear to have similar \lya\ and UV-continuum escape fractions (q $\sim$ 1), and some objects have \lya\ escape fractions $<$ the UV continuum (q $>$ 1).  \citet{scarlata09} recently investigated the ISMs of LAEs from the \citet{deharveng08} sample, examining a number of radiative transfer scenarios, including a clumpy ISM.  In particular, they study how the hydrogen line ratios change with the number of clumps, deducing that they have clumpy interstellar media.  However, these authors also conclude that there is no evidence for preferential shielding of \lya~from dust, as they do not have any objects with \lya/H$\alpha$ $>$ 8.7, similar to our results.  Scarlata et al.\ note that the attenuation curve in a clumpy ISM can differ from that for the uniform dust screen if the number of clumps are small.  Thus although we have assumed a large number of clumps such that the uniform dust screen law applies in our above discussion, this need not be the case.
\begin{deluxetable*}{cccccccc}
\tablecaption{Measured Dust Extinction and Gas--Phase Metallicities}\label{tab:final}
\tablewidth{0pt}
\tablehead{
\colhead{Object} & \colhead{E(B-V)$_{gas}$} & \colhead{A$_{V, stellar}$} & \colhead{Z$_{N2}$} & \colhead{Z$_{O3N2}$} & \colhead{Z$^{l}_{R23}$} & \colhead{Z$^{u}_{R23}$} & \colhead{Z Weighted Mean}\\
\colhead{$ $} & \colhead{$ $} & \colhead{(mag)} & \colhead{(Z\sol)} & \colhead{(Z\sol)} & \colhead{(Z\sol)} & \colhead{(Z\sol)} & \colhead{(Z\sol)}\\
}
\startdata
\phantom{E}EGS7&0.23 $\pm$ 0.02&0.41 $\pm$ 0.04&0.30 $\pm$ 0.13&0.27 $\pm$ 0.09&0.18 $\pm$ 0.03&0.68 $\pm$ 0.11&0.20 $\pm$ 0.03\\
\phantom{E}EGS8&0.06 $\pm$ 0.06&0.11 $\pm$ 0.11&0.26$^{\dagger}$&0.29$^{\dagger}$&0.20 $\pm$ 0.12&0.86 $\pm$ 0.36&0.25 $\pm$ 0.07\\
 EGS10&0.16 $\pm$ 0.04&0.29 $\pm$ 0.07&0.18$^{\dagger}$&0.23$^{\dagger}$&0.28 $\pm$ 0.11&0.66 $\pm$ 0.20&0.22 $\pm$ 0.05\\
EGS11&0.19 $\pm$ 0.07&0.33 $\pm$ 0.12&0.85 $\pm$ 0.38&0.73 $\pm$ 0.26&0.23 $\pm$ 0.17&1.11 $\pm$ 0.46&0.44 $\pm$ 0.13\\
EGS12&0.15 $\pm$ 0.05&0.27 $\pm$ 0.10&0.54 $\pm$ 0.24&0.48 $\pm$ 0.17&0.24 $\pm$ 0.13&0.90 $\pm$ 0.33&0.36 $\pm$ 0.10\\
EGS13&0.27 $\pm$ 0.02&0.47 $\pm$ 0.04&0.58 $\pm$ 0.24&0.53 $\pm$ 0.17&0.24 $\pm$ 0.05&0.96 $\pm$ 0.15&0.28 $\pm$ 0.05\\
EGS19&0.17 $\pm$ 0.05&0.30 $\pm$ 0.08&0.87 $\pm$ 0.38&0.81 $\pm$ 0.28&0.17 $\pm$ 0.09&1.31 $\pm$ 0.34&0.26 $\pm$ 0.08\\
EGS20&0.01 $\pm$ 0.04&0.02 $\pm$ 0.08&0.31$^{\dagger}$&0.31$^{\dagger}$&0.25 $\pm$ 0.11&0.73 $\pm$ 0.24&0.28 $\pm$ 0.07\\
EGS21&0.14 $\pm$ 0.09&0.26 $\pm$ 0.16&1.13 $\pm$ 0.53&1.33 $\pm$ 0.52&0.14 $\pm$ 0.14&1.74 $\pm$ 0.66&1.36 $\pm$ 0.33\\
EGS22&0.40 $\pm$ 0.11&0.72 $\pm$ 0.21&0.31$^{\dagger}$&0.52$^{\dagger}$&0.38 $\pm$ 0.59&1.07 $\pm$ 0.85&0.38 $\pm$ 0.13\\
EGS23&0.20 $\pm$ 0.02&0.36 $\pm$ 0.04&0.49 $\pm$ 0.21&0.41 $\pm$ 0.13&0.22 $\pm$ 0.05&0.80 $\pm$ 0.14&0.26 $\pm$ 0.05\\
EGS25&0.17 $\pm$ 0.01&0.31 $\pm$ 0.03&0.33 $\pm$ 0.14&0.28 $\pm$ 0.09&0.14 $\pm$ 0.02&0.73 $\pm$ 0.10&0.15 $\pm$ 0.02\\
\enddata
\tablecomments{$^{\dagger}$ [N\,{\sc ii}] was undetected in EGS8, EGS10, EGS20 and EGS22, and the 3 $\sigma$ upper limit was used on the line flux to compute N2 and O3N2.  N2 is thus an upper limit and O3N2 is a lower limit.  The weighted mean metallicity was computed with just one value for R23 per object, using the value which gave the lowest error on the weighted mean with the N2 and O3N2 metallicities (which ended up being the lower branch in all cases).  The exception is EGS21, where although the weighted mean uncertainty was formally smaller with the lower R23 result, the high metallicities with the O3N2 and N2 indices imply that the upper branch is correct.}
\end{deluxetable*}

Examining the results for all studied GALEX LAEs, it appears that although the ISMs in low--redshift analog LAEs span a wide range of extinction and clumpiness values, some level of clumping appears likely (i.e., a uniform dust screen does not appear to be supported by the data).  We note that the necessity of an aperture correction based on photometric data adds a level of overall uncertainty, as it assumes the H$\alpha$ flux is spatially correlated with the broadband emission.  This can be avoided in future studies when both lines are measured in similar apertures.

\section{Mass-Metallicity Relation}

Using our measured emission lines, we can measure the gas-phase metallicities of the galaxies in our sample with a number of metallicity indices.  We can then examine these galaxies on a mass-metallicity relation to see how they compare to other populations.  We use ratios of emission lines of oxygen and nitrogen to those of hydrogen to measure the gas--phase metallicities of the galaxies in our sample.  Specifically, we use the calibrations of the N2 and O3N2 indices of \citet{pettini04}, where 
\begin{equation}
\mathrm{N2 = log \left(\frac{[N II]}{H\alpha} \right)}
\end{equation}
and
\begin{equation}
\mathrm{O3N2 = log \left(\frac{[O III] \lambda 5007/H\beta}{[N II]/H\alpha} \right)}.
\end{equation}
Additionally, we obtain another measure of the metallicity using the well--known R23 index, where
\begin{equation}
R_{23} = \mathrm{\frac{[O II]\lambda 3727 + [O III]\lambda 4959,5007}{H\beta}},
\end{equation}
using the calibration of \citet{kobulnicky99}.  This index is double valued above $Z$ $\sim$ 0.3 Z\sol, thus we tabulated the metallicities for both the upper and lower branches.  Table 4 presents the measured metallicities of our galaxies from each of these indices, as well as their errors, including systematic uncertainties from the calibrations of 0.05, 0.14 and 0.18 dex for R23, O3N2 and N2 indices, respectively.  We note that these indices were calibrated using nearby galaxies, and they may be sensitive to a change in H\,{\sc ii} region physical conditions such as the ionization parameter or electron temperature.  In fact, many high-redshift star-forming galaxies may have an increased ionization parameter as they appear enhanced in [O\,{\sc iii}]/H$\beta$ from the local star-forming galaxies sequences from the SDSS \citep[e.g.,][]{erb06, finkelstein09d, hainline09}.  However, as shown in \citet{finkelstein09e} the star-forming members of our sample have similar [O\,{\sc iii}]/H$\beta$ ratios as the SDSS galaxies, thus they likely have similar H\,{\sc ii} region properties as well.

Prior to this computation, the emission lines were corrected for the inferred dust extinction from \S 4.  We also present a weighted mean metallicity from all three measures, where the R23 value which results in smaller uncertainty on the weighted mean was chosen (with the exception of EGS21, where it is obvious that the upper branch of R23 is the appropriate one).  In nearly all cases, the three measures of the metallicities are consistent within their uncertainties, thus our results are not dependent on the precise indices used.  Investigating Table 3, we find that these low--redshift LAEs typically have metallicities from 0.2 -- 0.4 Z\sol.  

\begin{figure*}
\epsscale{1}
\plotone{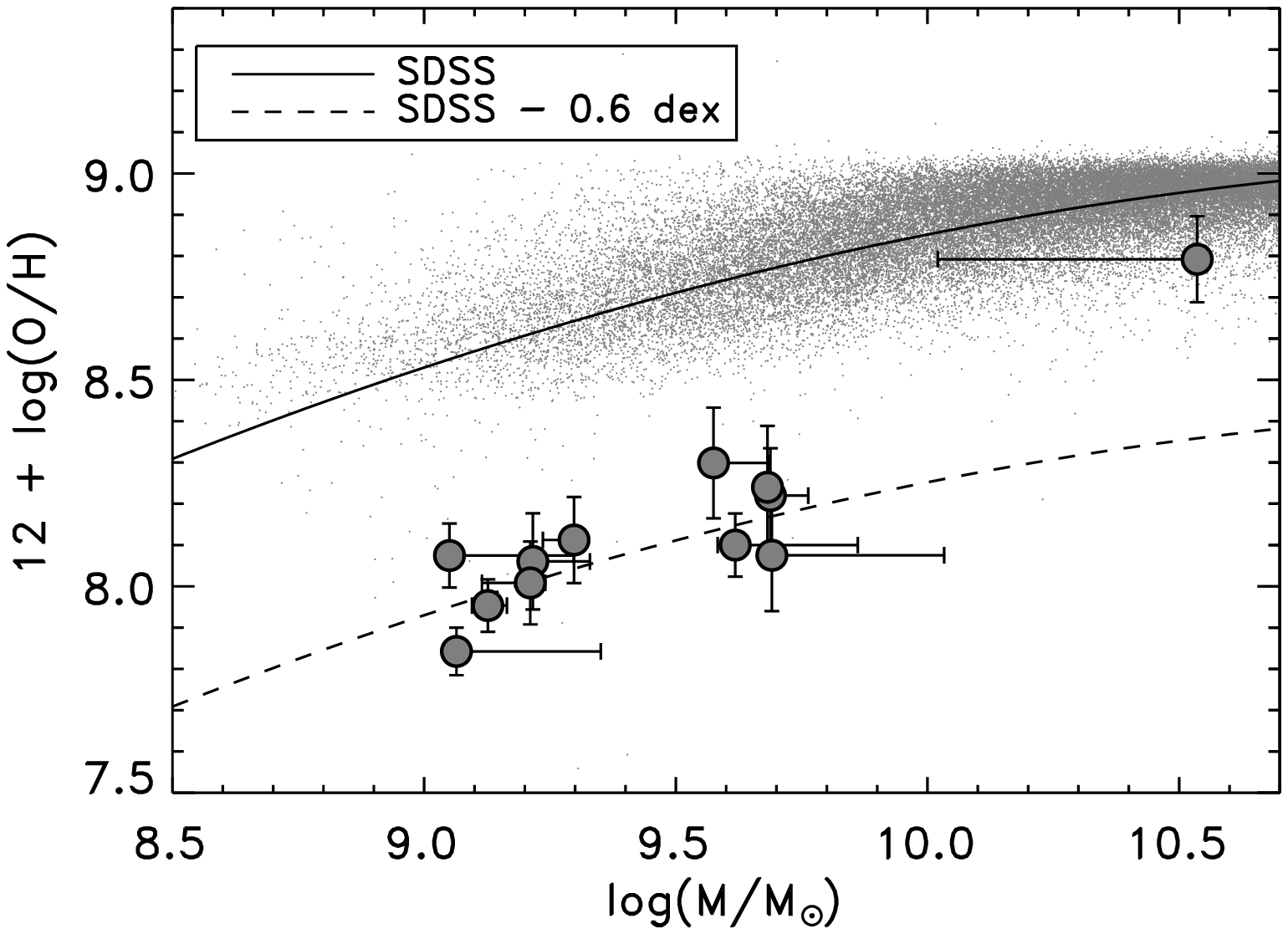}
\vspace{-4mm}
\caption{Our sample of 12 star--forming low--redshift LAE analogs plotted in a mass--metallicity plane.  For comparison, we show the sample of $\sim$ 53,000 SDSS galaxies from \citet{tremonti04}, where we recomputed their metallicities using the same methods as our sample.  The solid line is the best--fit mass--metallicity relation from \citet{tremonti04} (shifted down 0.1 dex to match the sequence shown using our metallicity derivation).  The dashed line is the SDSS relation shifted downward by $\sim$ 0.6 dex, highlighting that these LAE analogs are significantly shifted downward in metallicity for a given mass when compared to the SDSS sample.  This implies that galaxies exhibiting \lya\ in emission may be systematically more metal poor than other galaxies at the same stellar mass and redshift.}\label{fig:mz}
\end{figure*}

In \fig{fig:mz}, we plot the 12 star--forming z $\sim$ 0.3 LAEs in a mass--metallicity plane.  Star--forming galaxies were extensively studied in this plane by \citet{tremonti04}, who used SDSS spectra of $\sim$ 53,000 star--forming galaxies to study the mass--metallicity relation at z $\sim$ 0.1.  In that work, they computed the metallicities by using a combination of all strong emission lines.  Because different metallicity indicators suffer inherent systematic uncertainties of 0.2 -- 0.6 dex \citep{kewley02, tremonti04, moustakas10}, we recomputed metallicities for the SDSS sample using the same techniques as above.  In \fig{fig:mz} we show the best--fit mass--metallicity relation of \citet{tremonti04}, which is shifted downward by 0.1 dex to match our recomputed SDSS metallicities.

At any given mass, all but the most massive of our z $\sim$ 0.3 LAEs lie significantly below the z $\sim$ 0.1 mass--metallicity relation.  In \fig{fig:mz}, we show the best--fit curve to the z $\sim$ 0.1 mass--metallicity relation from \citet{tremonti04} shifted to match our derived SDSS metallicities, and we find that shifting it downward by a further $\sim$ 0.6 dex best matches our sample of LAEs.  This offset is larger than plausible errors in the stellar masses.  This illustrates that at low--redshift, galaxies exhibiting \lya~in emission have significantly lower metallicities than comparable mass galaxies with no such pre--selection.  \citet{cowie10} found similar results, as they noted that {\it GALEX}--selected LAEs had lower metallicities than star--forming galaxies selected via the near-UV continuum.  However, their brighter LAEs had similar metallicities to their continuum-selected sample, while all but one of our more massive LAEs are still more metal-poor than the SDSS sample.  The differences may be due to the use of solely the N2 index in \citet{cowie10}, as it typically yields the highest metallicities of the three indicators we use in our sample.

At $z \sim$ 0.3 the mass-metallicity relation of LAEs is shifted lower than that of the SDSS galaxies, which represent the general star-forming galaxy population at these redshifts.  At high redshift, LBGs appear to be the dominant galaxy population.  \citet{erb06} has studied the mass--metallicity relation of LBGs at z $\sim$ 2.3, finding that their objects fit the shape of the local mass--metallicity relation, but shifted downward in metallicity by 0.56 dex (over a metallicity range of 0.4 -- 0.9 Z\sol\ for galaxies with stellar masses of 10$^{9}$ -- 10$^{11}$ $\mathcal{M}$\sol).  Current results show that high--redshift LAEs are on average less evolved in age, mass and dust content than LBGs \citep[e.g,.][]{gawiser06a, finkelstein09a, pentericci09}, thus it may seem intuitive that they would be lower in metallicity as well (i.e., less evolved in all physical characteristics), possibly shifted even further down for the LBG mass-metallicity relation.

This was recently studied at high-redshift by \citet{finkelstein10d}, where upper limits on the metallicities of two LAEs at $z =$ 2.3 and 2.5 were measured to be $Z$ $<$ 0.17 and 0.28 $Z$\sol, respectively, with the N2 index.  The object with the lower metallicity upper limit appears to be significantly below even the $z =$ 2.3 LBGs of \citet{erb06} on the mass-metallicity relation, analogous to our results at $z \sim$ 0.3.  At $z =$ 2.3, LBGs represent the general star-forming galaxy population, and LAEs appear to be more metal-poor than this population (though with a sample of two objects we cannot draw robust conclusions).  However, we see something very similar at $z \sim$ 0.3, where the SDSS sample represents the general star-forming galaxy population.  Thus, it may be that galaxies exhibiting \lya\ in emission have preferentially lower metallicities.  However, the evidence for this at high redshift is not strong, and a confident resolution will not be possible until we can measure the metallicities of large samples of high--redshift LAEs, which will be done in the near future with the upcoming multi-object NIR spectrographs.

\section{Conclusions}

We present results from a spectroscopic survey of {\it GALEX}--discovered LAEs at z $\sim$ 0.3.  These objects are useful as nearby analogs to high--redshift LAEs, as we are able to directly measure physical quantities which are only inferred at high--redshift.  In this Paper, we present the full optical spectra of 12 star-forming LAEs, including hydrogen, oxygen and nitrogen line fluxes.  We first re-examine their stellar population properties, and find that when excluding the recently identified AGNs in the parent sample, the average ages and masses of the remaining star-forming LAEs are lower, at $\sim$ 300 Myr and 4 $\times$ 10$^{10}$ $\mathcal{M}$\sol\, respectively.  These are older and more massive than a typical high-redshift LAE, but similar to a rare population of evolved high-redshift LAEs.

Correcting our measured emission line fluxes for stellar absorption, we use the Balmer decrement to investigate the level of dust attenuation in these galaxies.  We find that these galaxies all contain some level of dust, with an average extinction of A$_\mathrm{V}$ $\sim$ 0.3 mag.  This is similar to what is typically found at high redshift, and is also consistent with the overall trend of the existence of dust in galaxies which are still able to emit in \lya.  Using the ratio of \lya/H$\alpha$ and the measured extinction as a probe of the ISM geometry, we find that in general the ISM geometries for the majority of the sample appear to be somewhere between homogeneous and purely clumpy, implying that a uniform dust screen geometry does not apply to these objects.

Lastly, we use three different metallicity indicators to measure the gas--phase metallicities of the LAEs in our sample.  We find that all but one of these galaxies are have lower metallicities than field galaxies at the same stellar mass, with typical gas--phase metallicities of $\sim$ 0.2 -- 0.4 Z\sol.  Comparing to recent results at high-redshift, it appears that galaxies exhibiting \lya\ in emission may be systematically more metal poor at all redshifts, though more work is needed.

Studying LAEs spectroscopically allows us to learn about their physical properties in more detail and with greater confidence than with stellar population model fitting.  Ideally, these low--redshift LAEs would be used as analogs to allow us to decipher the physical characteristics of high--redshift LAEs.  As these objects studied here are more evolved that typical narrowband--selected high--redshift LAEs, we can only put limits on what we expect to find at high--redshift.  However, with the numerous multi--object near--infrared spectrographs coming online soon, we can expect to make great progress in understanding LAEs by directly measuring their physical properties at z $\sim$ 2 -- 3.

\acknowledgements

We thank the anonymous referee for an extremely useful report which improved the quality of this paper.  SLF is supported by the George P. and Cynthia Woods Mitchell Institute for Fundamental Physics and Astronomy in the Department of Physics and Astronomy at Texas A\&M University.


\begin{thebibliography}{47}
\expandafter\ifx\csname natexlab\endcsname\relax\def\natexlab#1{#1}\fi

\bibitem[{{Atek} {et~al.}(2008){Atek}, {Kunth}, {Hayes}, {{\"O}stlin}, \&
  {Mas-Hesse}}]{atek08}
{Atek}, H., {Kunth}, D., {Hayes}, M., {{\"O}stlin}, G., \& {Mas-Hesse}, J.~M.
  2008, \aap, 488, 491

\bibitem[{{Atek} {et~al.}(2009){Atek}, {Kunth}, {Schaerer}, {Hayes},
  {Deharveng}, {{\"O}stlin}, \& {Mas-Hesse}}]{atek09}
{Atek}, H., {Kunth}, D., {Schaerer}, D., {Hayes}, M., {Deharveng}, J.~M.,
  {{\"O}stlin}, G., \& {Mas-Hesse}, J.~M. 2009, \aap, 506, L1

\bibitem[{{Bertin} \& {Arnouts}(1996)}]{bertin96}
{Bertin}, E., \& {Arnouts}, S. 1996, \aaps, 117, 393

\bibitem[{{Blanc} {et~al.}(2010){Blanc}, {Adams}, {Gebhardt}, {Hill}, {Drory},
  {Hao}, {Bender}, {Ciardullo}, {Finkelstein}, \& {Gawiser}}]{blanc10}
{Blanc}, G.~A., {et~al.} 2010, ApJ Submitted

\bibitem[{{Bruzual} \& {Charlot}(2003)}]{bruzual03}
{Bruzual}, G., \& {Charlot}, S. 2003, \mnras, 344, 1000

\bibitem[{{Calzetti} {et~al.}(2000){Calzetti}, {Armus}, {Bohlin}, {Kinney},
  {Koornneef}, \& {Storchi-Bergmann}}]{calzetti00}
{Calzetti}, D., {Armus}, L., {Bohlin}, R.~C., {Kinney}, A.~L., {Koornneef}, J.,
  \& {Storchi-Bergmann}, T. 2000, \apj, 533, 682

\bibitem[{{Cardelli} {et~al.}(1989){Cardelli}, {Clayton}, \&
  {Mathis}}]{cardelli89}
{Cardelli}, J.~A., {Clayton}, G.~C., \& {Mathis}, J.~S. 1989, \apj, 345, 245

\bibitem[{{Chabrier}(2003)}]{chabrier03}
{Chabrier}, G. 2003, \pasp, 115, 763

\bibitem[{{Cowie} {et~al.}(2010){Cowie}, {Barger}, \& {Hu}}]{cowie10}
{Cowie}, L.~L., {Barger}, A.~J., \& {Hu}, E.~M. 2010, \apj, 711, 928

\bibitem[{{Deharveng} {et~al.}(2008){Deharveng}, {Small}, {Barlow},
  {P{\'e}roux}, {Milliard}, {Friedman}, {Martin}, {Morrissey}, {Schiminovich},
  {Forster}, {Seibert}, {Wyder}, {Bianchi}, {Donas}, {Heckman}, {Lee},
  {Madore}, {Neff}, {Rich}, {Szalay}, {Welsh}, \& {Yi}}]{deharveng08}
{Deharveng}, J.-M., {et~al.} 2008, \apj, 680, 1072

\bibitem[{{Erb} {et~al.}(2006{\natexlab{a}}){Erb}, {Shapley}, {Pettini},
  {Steidel}, {Reddy}, \& {Adelberger}}]{erb06b}
{Erb}, D.~K., {Shapley}, A.~E., {Pettini}, M., {Steidel}, C.~C., {Reddy},
  N.~A., \& {Adelberger}, K.~L. 2006{\natexlab{a}}, \apj, 644, 813

\bibitem[{{Erb} {et~al.}(2006{\natexlab{b}}){Erb}, {Steidel}, {Shapley},
  {Pettini}, {Reddy}, \& {Adelberger}}]{erb06}
{Erb}, D.~K., {Steidel}, C.~C., {Shapley}, A.~E., {Pettini}, M., {Reddy},
  N.~A., \& {Adelberger}, K.~L. 2006{\natexlab{b}}, \apj, 646, 107

\bibitem[{{Fabricant} {et~al.}(2005){Fabricant}, {Fata}, {Roll}, {Hertz},
  {Caldwell}, {Gauron}, {Geary}, {McLeod}, {Szentgyorgyi}, {Zajac}, {Kurtz},
  {Barberis}, {Bergner}, {Brown}, {Conroy}, {Eng}, {Geller}, {Goddard},
  {Honsa}, {Mueller}, {Mink}, {Ordway}, {Tokarz}, {Woods}, {Wyatt}, {Epps}, \&
  {Dell'Antonio}}]{fabricant05}
{Fabricant}, D., {et~al.} 2005, \pasp, 117, 1411

\bibitem[{{Finkelstein} {et~al.}(2009{\natexlab{a}}){Finkelstein}, {Cohen},
  {Malhotra}, \& {Rhoads}}]{finkelstein09c}
{Finkelstein}, S.~L., {Cohen}, S.~H., {Malhotra}, S., \& {Rhoads}, J.~E.
  2009{\natexlab{a}}, \apj, 700, 276

\bibitem[{{Finkelstein} {et~al.}(2009{\natexlab{b}}){Finkelstein}, {Cohen},
  {Malhotra}, {Rhoads}, {Papovich}, {Zheng}, \& {Wang}}]{finkelstein09e}
{Finkelstein}, S.~L., {Cohen}, S.~H., {Malhotra}, S., {Rhoads}, J.~E.,
  {Papovich}, C., {Zheng}, Z.~Y., \& {Wang}, J. 2009{\natexlab{b}}, \apjl, 703,
  L162

\bibitem[{{Finkelstein} {et~al.}(2009{\natexlab{c}}){Finkelstein}, {Papovich},
  {Rudnick}, {Egami}, {Le Floc'h}, {Rieke}, {Rigby}, \&
  {Willmer}}]{finkelstein09d}
{Finkelstein}, S.~L., {Papovich}, C., {Rudnick}, G., {Egami}, E., {Le Floc'h},
  E., {Rieke}, M.~J., {Rigby}, J.~R., \& {Willmer}, C.~N.~A.
  2009{\natexlab{c}}, \apj, 700, 376

\bibitem[{{Finkelstein} {et~al.}(2009{\natexlab{d}}){Finkelstein}, {Rhoads},
  {Malhotra}, \& {Grogin}}]{finkelstein09a}
{Finkelstein}, S.~L., {Rhoads}, J.~E., {Malhotra}, S., \& {Grogin}, N.
  2009{\natexlab{d}}, \apj, 691, 465

\bibitem[{{Finkelstein} {et~al.}(2008){Finkelstein}, {Rhoads}, {Malhotra},
  {Grogin}, \& {Wang}}]{finkelstein08}
{Finkelstein}, S.~L., {Rhoads}, J.~E., {Malhotra}, S., {Grogin}, N., \& {Wang},
  J. 2008, \apj, 678, 655

\bibitem[{{Finkelstein} {et~al.}(2007){Finkelstein}, {Rhoads}, {Malhotra},
  {Pirzkal}, \& {Wang}}]{finkelstein07}
{Finkelstein}, S.~L., {Rhoads}, J.~E., {Malhotra}, S., {Pirzkal}, N., \&
  {Wang}, J. 2007, \apj, 660, 1023

\bibitem[{{Finkelstein} {et~al.}(2010){Finkelstein}, {Hill}, {Gebhardt},
  {Adams}, {Blanc}, {Papovich}, {Ciardullo}, {Drory}, {Gawiser}, {Gronwall},
  {Schneider}, \& {Tran}}]{finkelstein10d}
{Finkelstein}, S.~L., {et~al.} 2010, ApJ Submitted, astroph/1011.0431

\bibitem[{{Gawiser} {et~al.}(2006){Gawiser}, {van Dokkum}, {Herrera}, {Maza},
  {Castander}, {Infante}, {Lira}, {Quadri}, {Toner}, {Treister}, {Urry},
  {Altmann}, {Assef}, {Christlein}, {Coppi}, {Dur{\'a}n}, {Franx}, {Galaz},
  {Huerta}, {Liu}, {L{\'o}pez}, {M{\'e}ndez}, {Moore}, {Rubio}, {Ruiz}, {Toft},
  \& {Yi}}]{gawiser06a}
{Gawiser}, E., {et~al.} 2006, \apjs, 162, 1

\bibitem[{{Hainline} {et~al.}(2009){Hainline}, {Shapley}, {Kornei}, {Pettini},
  {Buckley-Geer}, {Allam}, \& {Tucker}}]{hainline09}
{Hainline}, K.~N., {Shapley}, A.~E., {Kornei}, K.~A., {Pettini}, M.,
  {Buckley-Geer}, E., {Allam}, S.~S., \& {Tucker}, D.~L. 2009, \apj, 701, 52

\bibitem[{{Hansen} \& {Oh}(2006)}]{hansen06}
{Hansen}, M., \& {Oh}, S.~P. 2006, \mnras, 367, 979

\bibitem[{{Hayes} {et~al.}(2007){Hayes}, {{\"O}stlin}, {Atek}, {Kunth},
  {Mas-Hesse}, {Leitherer}, {Jim{\'e}nez-Bail{\'o}n}, \& {Adamo}}]{hayes07}
{Hayes}, M., {{\"O}stlin}, G., {Atek}, H., {Kunth}, D., {Mas-Hesse}, J.~M.,
  {Leitherer}, C., {Jim{\'e}nez-Bail{\'o}n}, E., \& {Adamo}, A. 2007, \mnras,
  382, 1465

\bibitem[{{Kauffmann} {et~al.}(2003){Kauffmann}, {Heckman}, {White}, {Charlot},
  {Tremonti}, {Brinchmann}, {Bruzual}, {Peng}, {Seibert}, {Bernardi},
  {Blanton}, {Brinkmann}, {Castander}, {Cs{\'a}bai}, {Fukugita}, {Ivezic},
  {Munn}, {Nichol}, {Padmanabhan}, {Thakar}, {Weinberg}, \&
  {York}}]{kauffmann03a}
{Kauffmann}, G., {et~al.} 2003, \mnras, 341, 33

\bibitem[{{Kewley} \& {Dopita}(2002)}]{kewley02}
{Kewley}, L.~J., \& {Dopita}, M.~A. 2002, \apjs, 142, 35

\bibitem[{{Kobulnicky} {et~al.}(1999){Kobulnicky}, {Kennicutt}, \&
  {Pizagno}}]{kobulnicky99}
{Kobulnicky}, H.~A., {Kennicutt}, Jr., R.~C., \& {Pizagno}, J.~L. 1999, \apj,
  514, 544

\bibitem[{{Kornei} {et~al.}(2010){Kornei}, {Shapley}, {Erb}, {Steidel},
  {Reddy}, {Pettini}, \& {Bogosavljevi{\'c}}}]{kornei10}
{Kornei}, K.~A., {Shapley}, A.~E., {Erb}, D.~K., {Steidel}, C.~C., {Reddy},
  N.~A., {Pettini}, M., \& {Bogosavljevi{\'c}}, M. 2010, \apj, 711, 693

\bibitem[{{Kudritzki} {et~al.}(2000){Kudritzki}, {M{\'e}ndez}, {Feldmeier},
  {Ciardullo}, {Jacoby}, {Freeman}, {Arnaboldi}, {Capaccioli}, {Gerhard}, \&
  {Ford}}]{kudritzki00}
{Kudritzki}, R., {et~al.} 2000, \apj, 536, 19

\bibitem[{{Kunth} {et~al.}(1998){Kunth}, {Mas-Hesse}, {Terlevich}, {Terlevich},
  {Lequeux}, \& {Fall}}]{kunth98}
{Kunth}, D., {Mas-Hesse}, J.~M., {Terlevich}, E., {Terlevich}, R., {Lequeux},
  J., \& {Fall}, S.~M. 1998, \aap, 334, 11

\bibitem[{{Lai} {et~al.}(2007){Lai}, {Huang}, {Fazio}, {Cowie}, {Hu}, \&
  {Kakazu}}]{lai07}
{Lai}, K., {Huang}, J.-S., {Fazio}, G., {Cowie}, L.~L., {Hu}, E.~M., \&
  {Kakazu}, Y. 2007, \apj, 655, 704

\bibitem[{{Lai} {et~al.}(2008){Lai}, {Huang}, {Fazio}, {Gawiser}, {Ciardullo},
  {Damen}, {Franx}, {Gronwall}, {Labb{\'e}}, {Magdis}, \& {van Dokkum}}]{lai08}
{Lai}, K., {et~al.} 2008, \apj, 674, 70

\bibitem[{{Malhotra} \& {Rhoads}(2002)}]{malhotra02}
{Malhotra}, S., \& {Rhoads}, J.~E. 2002, \apjl, 565, L71

\bibitem[{{Morrissey} {et~al.}(2007){Morrissey}, {Conrow}, {Barlow}, {Small},
  {Seibert}, {Wyder}, {Budav{\'a}ri}, {Arnouts}, {Friedman}, {Forster},
  {Martin}, {Neff}, {Schiminovich}, {Bianchi}, {Donas}, {Heckman}, {Lee},
  {Madore}, {Milliard}, {Rich}, {Szalay}, {Welsh}, \& {Yi}}]{morrissey07}
{Morrissey}, P., {et~al.} 2007, \apjs, 173, 682

\bibitem[{{Moustakas} {et~al.}(2010){Moustakas}, {Kennicutt}, {Tremonti},
  {Dale}, {Smith}, \& {Calzetti}}]{moustakas10}
{Moustakas}, J., {Kennicutt}, Jr., R.~C., {Tremonti}, C.~A., {Dale}, D.~A.,
  {Smith}, J., \& {Calzetti}, D. 2010, \apjs, 190, 233

\bibitem[{{Neufeld}(1991)}]{neufeld91}
{Neufeld}, D.~A. 1991, \apjl, 370, L85

\bibitem[{{Nilsson} {et~al.}(2007){Nilsson}, {M{\o}ller}, {M{\"o}ller},
  {Fynbo}, {Micha{\l}owski}, {Watson}, {Ledoux}, {Rosati}, {Pedersen}, \&
  {Grove}}]{nilsson07a}
{Nilsson}, K.~K., {et~al.} 2007, \aap, 471, 71

\bibitem[{{Osterbrock}(1989)}]{osterbrock89}
{Osterbrock}, D.~E. 1989, {Astrophysics of Gaseous Nebulae and Active Galactic
  Nuclei} (Mill Valley: University Science Books)

\bibitem[{{{\"O}stlin} {et~al.}(2009){{\"O}stlin}, {Hayes}, {Kunth},
  {Mas-Hesse}, {Leitherer}, {Petrosian}, \& {Atek}}]{ostlin09}
{{\"O}stlin}, G., {Hayes}, M., {Kunth}, D., {Mas-Hesse}, J.~M., {Leitherer},
  C., {Petrosian}, A., \& {Atek}, H. 2009, \aj, 138, 923

\bibitem[{{Partridge} \& {Peebles}(1967)}]{partridge67}
{Partridge}, R.~B., \& {Peebles}, P.~J.~E. 1967, \apj, 147, 868

\bibitem[{{Pentericci} {et~al.}(2009){Pentericci}, {Grazian}, {Fontana},
  {Castellano}, {Giallongo}, {Salimbeni}, \& {Santini}}]{pentericci09}
{Pentericci}, L., {Grazian}, A., {Fontana}, A., {Castellano}, M., {Giallongo},
  E., {Salimbeni}, S., \& {Santini}, P. 2009, \aap, 494, 553

\bibitem[{{Pettini} \& {Pagel}(2004)}]{pettini04}
{Pettini}, M., \& {Pagel}, B.~E.~J. 2004, \mnras, 348, L59

\bibitem[{{Pirzkal} {et~al.}(2007){Pirzkal}, {Malhotra}, {Rhoads}, \&
  {Xu}}]{pirzkal07}
{Pirzkal}, N., {Malhotra}, S., {Rhoads}, J.~E., \& {Xu}, C. 2007, \apj, 667, 49

\bibitem[{{Scarlata} {et~al.}(2009){Scarlata}, {Colbert}, {Teplitz}, {Panagia},
  {Hayes}, {Siana}, {Rau}, {Francis}, {Caon}, {Pizzella}, \&
  {Bridge}}]{scarlata09}
{Scarlata}, C., {et~al.} 2009, \apjl, 704, L98

\bibitem[{{Storey} \& {Zeippen}(2000)}]{storey00}
{Storey}, P.~J., \& {Zeippen}, C.~J. 2000, \mnras, 312, 813

\bibitem[{{Tremonti} {et~al.}(2004){Tremonti}, {Heckman}, {Kauffmann},
  {Brinchmann}, {Charlot}, {White}, {Seibert}, {Peng}, {Schlegel}, {Uomoto},
  {Fukugita}, \& {Brinkmann}}]{tremonti04}
{Tremonti}, C.~A., {et~al.} 2004, \apj, 613, 898

\bibitem[{{Walcher} {et~al.}(2010){Walcher}, {Groves}, {Budav{\'a}ri}, \&
  {Dale}}]{walcher10}
{Walcher}, J., {Groves}, B., {Budav{\'a}ri}, T., \& {Dale}, D. 2010, \apss, 257

\end{thebibliography}
\end{document}